\documentclass[a4paper,11pt]{article}
\pdfoutput=1 

\usepackage{jheppub} 

\usepackage[T1]{fontenc} 

\usepackage{graphicx}
\usepackage{color}
\usepackage{bm}
\usepackage{amsmath}
\usepackage{amsfonts}
\usepackage{amssymb}
\usepackage{mathrsfs}
\usepackage{enumerate}

\def\be{\begin{equation}}
\def\ee{\end{equation}}
\def\eq#1{{Eq.~(\ref{#1})}}
\def\eqs#1{{Eqs.~(\ref{#1})}}
\def\sect#1{{Sec.~\ref{#1}}}
\def\fig#1{{Fig.~\ref{#1}}}

\title{A quantum peek inside the black hole event horizon}

\author[a]{Sumanta Chakraborty,}
\author[a,b]{Suprit Singh}
\author[a]{and T. Padmanabhan}

\affiliation[a]{Inter-University Centre for Astronomy and Astrophysics,\\Ganeshkhind, Pune 411 007, India}
\affiliation[b]{Department of Physics and Astrophysics,\\University of Delhi, New Delhi 110 007, India}

\emailAdd{sumanta@iucaa.ernet.in}
\emailAdd{suprit@iucaa.ernet.in}
\emailAdd{paddy@iucaa.ernet.in}

\abstract{We solve the Klein-Gordon equation for a scalar field, in the background geometry of a dust cloud collapsing to form a black hole, everywhere in the (1+1) spacetime: that is, both inside and outside the event horizon and arbitrarily close to the curvature singularity. This allows us to determine the regularized stress tensor expectation value, everywhere in the appropriate quantum state (viz., the Unruh vacuum) of the field. We use this to study the behaviour of energy density and the flux measured in local inertial frames for the radially freely falling observer at any given event. Outside the black hole, energy density and flux lead to the standard results expected from the Hawking radiation emanating from the black hole, as the collapse proceeds. Inside the collapsing dust ball, the energy densities of both matter and scalar field diverge near the singularity in both (1+1) and (1+3) spacetime dimensions; but the energy density of the field dominates over that of classical matter. In the (1+3) 
dimensions, the total energy (of both scalar field and classical matter) inside a small spatial volume around the singularity is finite (and goes to zero as the size of the region goes to zero) but the total energy of the quantum field still dominates over that of the classical matter. Inside the event horizon, but \textit{outside} the collapsing matter, freely falling observers find that the energy density and the flux diverge close to the singularity. In this region, even the  integrated energy inside a small spatial volume enclosing the singularity  diverges. This result holds in both (1+1) and (1+3) spacetime dimensions  with a \emph{milder} divergence for the total energy inside a small region in (1+3) dimensions. These results suggest that the back-reaction effects are significant even in the region \emph{outside the matter but inside the event horizon}, close to the singularity.}

\begin{document}
\maketitle
\flushbottom

\section{Introduction, Motivation and Summary of Results}\label{Sec01n}

It is well known that, in the presence of a gravitationally collapsing structure forming a black hole and a quantum field in the Unruh vacuum state, an observer far away from the black hole will see a flux of thermal radiation at late times \cite{Hawking1974,Hawking1975}. This   result, which arises from the study of quantum fields in the curved spacetime, has led to several fascinating developments (see e.g. the textbooks and the reviews, \cite{Birrel1982,Padmanabhan2005a,Padmanabhan2010a, Helfer2003, Fabbri2005, Mukhanov2007, Parker2009, Padmanabhan2010b, Visser2003, Takagi1986}) in general relativity. While probing this result from different perspectives, quantum field theory in the region \textit{outside}  the black hole event horizon has been studied extensively in the literature. However, somewhat surprisingly, there has been much less emphasis in the study of quantum field theory \textit{inside} the event horizon (for some earlier work, similar in spirit, see e.g., \cite{Ellis2015,Ellis2014,Ellis2013,Hamilton2010a,Hamilton2010b,Aseem2009}). The purpose of this paper is to investigate quantum field theory inside the horizon in the context of a collapsing dust sphere in $(1+1)$ spacetime. As we shall see, such a study leads to several curious and interesting results which should have their counterparts in (1+3) dimensions.

\begin{figure}[t!]
\centering
\includegraphics[scale=1]{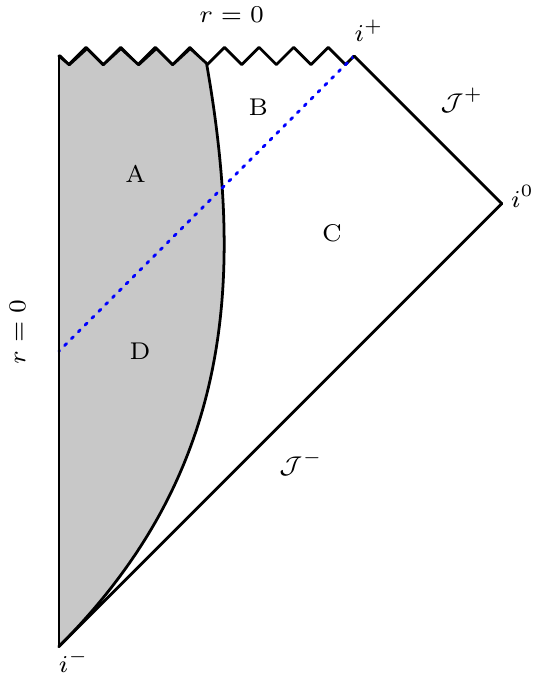}\hspace{2.0cm}
\includegraphics[scale=1]{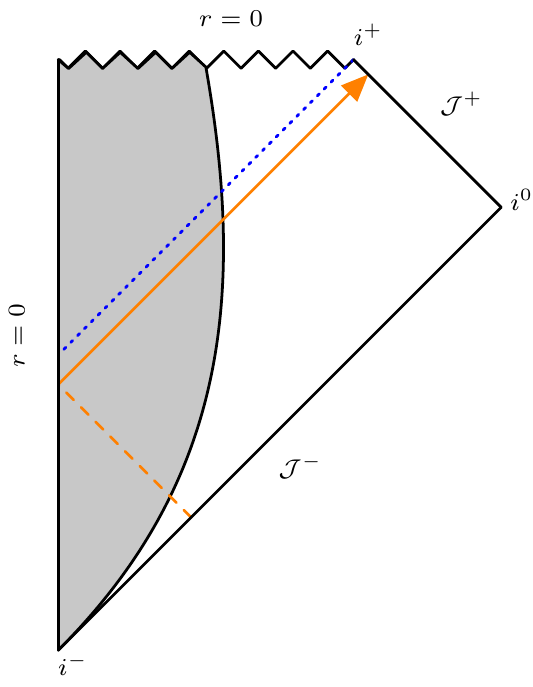}\\
\includegraphics[scale=1 ]{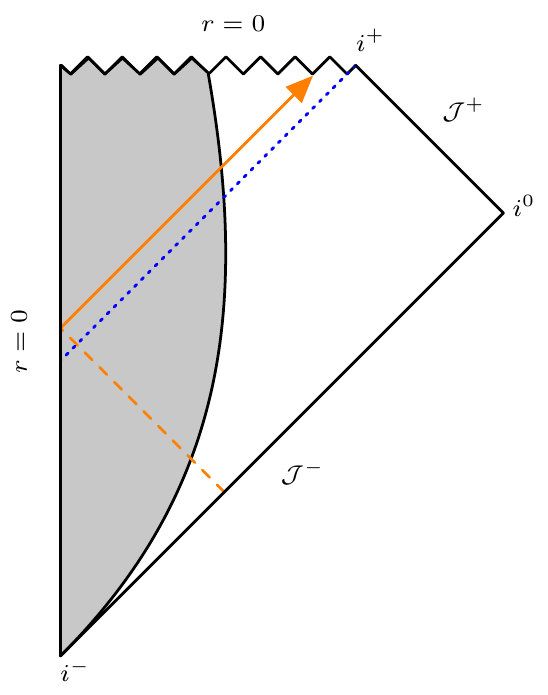}\hspace{2.0cm}
\includegraphics[scale=1]{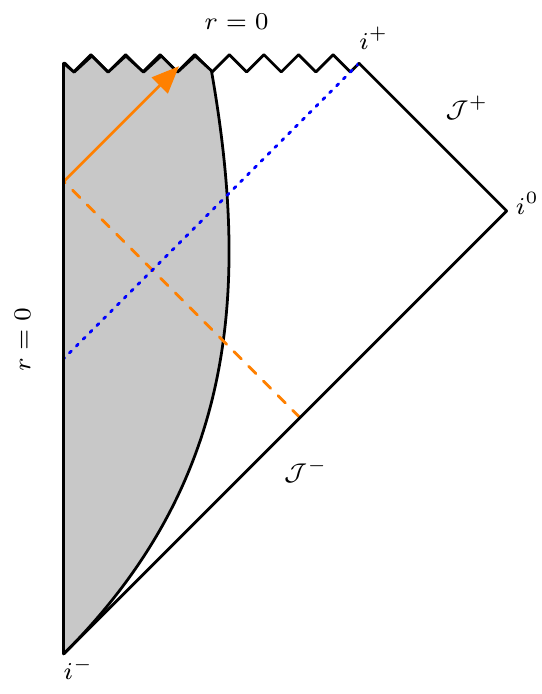}
\caption{The  Penrose diagram in the top left figure illustrates the four spacetime regions A,B,C,D we are interested in. The other three figures describe the three null rays relevant to this work. See text for detailed discussion.}
\label{Penrose}
\end{figure}

In order to elaborate what is involved, let us consider the Penrose diagram in \fig{Penrose} (top left) describing a gravitationally collapsing body. It is clear from the figure  that there are four distinct spacetime regions -- marked A, B, C and D. Of these, region D --- which is inside the collapsing body and outside the event horizon --- is the least interesting one for our purposes. Even though the time dependence of the metric will lead to particle production in this region, we do not expect any universal behaviour here; the results will depend on the details of the collapse. Let us next consider region C which is outside both the collapsing body and  the event horizon. This region is of primary importance --- and has been extensively investigated in the literature --- in connection with the black hole radiation. This is schematically illustrated  in \fig{Penrose} (top right) by  an outgoing null ray that straddles just outside the horizon and escapes to future null infinity. The thermal nature of the black hole radiation arises essentially due to the exponential redshift suffered by  this null ray as it travels from just outside the collapsing matter to future null infinity. While this ray is inside the collapsing matter during part of its travel, the details of the collapse are sub-dominant to the effect of the exponential redshift at late times. We can investigate the black hole evaporation scenario vis-a-vis different kinds of observers in this region: like, e.g,  asymptotic and non-asymptotic static observers, radial and inspiraling free-fallers,   observers moving in circular orbits etc.; all these cases indeed have been studied in the literature (see, for some recent work, Refs. \cite{Suprit2014b, Suprit2014a, Smerlak2013} which contain references to earlier papers). In this paper too, we will briefly discuss the physics in this region since recovering the standard results provides a `calibration test' for our approach and calculations.

But what we will concentrate on are the regions B and A which are \emph{inside} the event horizon. (We have not found any extensive and systematic investigation of these regions in published literature which was one of the key motivations for this work.) The examination of the Penrose diagram in \fig{Penrose} reveals the following facts about these two regions. 

\begin{itemize}

\item Region B is inside the event horizon but \emph{outside} the collapsing body. 
Being a vacuum region in a spherically symmetric geometry, this region is indeed described by a Schwarzschild metric. But, if we use the standard $(t, r, \theta, \phi)$ Schwarzschild coordinates, then $r$ is like a  \textit{time} coordinate in this region due to the flip of signs in the metric coefficients at $r< 2M$. Naively speaking, this makes the geometry ``time dependent'' (due to the dependence of geometry on $r$) in this region. Alternatively, one can describe this region using a coordinate system which is non-singular at the event horizon like, for e.g., the Kruskal coordinates, in which the line element takes the following form:
\begin{align}
ds^{2}=\frac{32M^{3}}{r}e^{-r/2M}\left(-dT^{2}+dX^{2}\right)+dL_{\perp}^{2}
\end{align}
where, $r$ is given as an implicit function of $X$ and $T$ via the relation: $(r/2M -1)e^{r/2M}=X^{2}-T^{2}$ and the transverse line element is $dL_{\perp}^{2}=r^{2}d\theta ^{2}+r^{2}sin ^{2}\theta d\phi ^{2}$. Once again, the metric will be time dependent because of its dependence in the Kruskal time coordinate $T$.  All these suggest that we expect a non-trivial dynamics for the quantum field in the region B.

A typical null ray in this region,  shown in \fig{Penrose} (bottom left), is trapped and hits the singularity \textit{in the region outside the collapsing matter} in a finite time. (More precisely, it gets \emph{arbitrarily} close to the singularity in the course of time.) It is instructive to compare the null ray in this region with the null ray in the region C mentioned above. (That is, we compare the figures at top right and bottom left of \fig{Penrose}.) The two relevant null rays propagate straddling the event horizon just  outside and just inside. While the outside ray  reaches future null infinity (and plays a key role in the description of the black hole evaporation), it is not clear how the physical parameters vary along the ray which travels just inside the event horizon but gets \textit{arbitrarily} close to the singularity at late times. This is important because the accumulation of energy density due to these modes of the vacuum \textit{which are trapped inside the event horizon} can have 
important back reaction effects. One key aim of this work is to study the physical properties of the quantum field at the events along the null ray like the one shown in the bottom left figure of \fig{Penrose}.

\item Region A is inside the collapsing body as well as inside the event horizon. Here too we can study the behaviour of the quantum field along the events in the path of the null ray shown in bottom right of the \fig{Penrose}. This null ray also hits the singularity but inside the collapsing matter. The key difference between the events along the rays shown in bottom right and those in bottom left, is the following: In the former case, one could compare the energy density of the quantum field with that of the collapsing matter; We expect both to diverge as we approach the singularity inside the collapsing matter. The key question is to determine which one diverges faster in order to ascertain the effects of back reaction. But in the latter case (corresponding to the situation in bottom left figure), we have no matter energy density to compare with the energy density of quantum fields close to the singularity. Therefore, any divergence in the energy density of the quantum field will potentially have 
important consequences for the back reaction. 

\end{itemize}     
We shall focus mostly on the regions A and B in this paper. We will use C mainly for ``calibration"; that is, to arrange the quantum state in such a way that an asymptotic observer sees the standard Hawking flux at late times. After choosing such a state, we will study the dynamical evolution  of the  quantum field and its regularized energy momentum tensor in the regions A and B. The region D is of comparatively less interest but we mention the results regarding this region in the appendix for the sake of completeness.  

\paragraph*{Methodology of the paper.} The Klein-Gordon equation for a massless scalar field can be solved exactly in $(1+1)$ dimension, by exploiting the conformal invariance of the theory, which is a fairly standard procedure (see for a review \cite{Brout1995}).  The conformal invariance of the massless scalar field in $(1+1)$ dimension (and the fact that any $(1+1)$ dimensional metric can be written in conformally flat form) makes the relevant mode functions just plane waves. Therefore, the dynamics of a massless scalar field can be reduced essentially to ray tracing. As argued usually in the literature \cite{Fabbri2005,Brout1995}, we expect the main conclusions to carry forward to $(1+3)$ dimensions, because the dominant contribution to the Hawking effect comes from $s$-waves even in $(1+3)$ dimensions.

Once the solution $\phi(x)$ to the Klein-Gordon equation is known (with suitable boundary conditions ensuring that one obtains the standard black hole evaporation in region C of the spacetime), our next task is to construct physically meaningful observables. There are two standard approaches which have been pursued in the literature in this context. One possibility is to introduce the particle detectors \cite{Unruh1976, Langlois2006,Louko2008,Satz2007} in the spacetime moving on various  trajectories. The particle content determined by the detector is then given essentially by the Fourier transform  of the  two-point correlation function of the field in the relevant quantum state. This leads to a Planckian spectrum for an asymptotic detector at rest in the Schwarzschild spacetime which agrees with the standard interpretation of black hole evaporation. Unfortunately, the response of the detector essentially measures the nature of vacuum fluctuations and is sensitive to the history of the trajectory because it 
is defined using an integral over the proper time. We cannot use the particle content determined by such a detector for estimating the back reaction effects of the quantum field on the spacetime. This is easily seen from the fact that a uniformly accelerated detector in flat spacetime will detect a thermal spectrum of particles but these ``particles'' do not back react on the spacetime in a generally covariant manner.

Since our primary interest is to study the effect of quantum fields on the background geometry, we need to use a more covariant diagnostic. Such a diagnostic is provided by the  (regularised) stress-energy tensor of the quantum field in the given quantum state. It is generally believed that, at least in the semi-classical regime, this regularized expectation value $\langle T_{ab}\rangle_{\rm reg}$ can be used as a source in Einstein's equation to study the effects of back reaction. In particular, when $\langle T_{ab}\rangle_{\rm reg}$ is comparable to the classical source of geometry $T_{ab}$, we will expect back reaction effects to be significant.

It should be noted that these two diagnostics for describing the quantum field --- viz., the detector response or $\langle T_{ab}\rangle_{\rm reg}$ --- will, in  general, give different results. Again, in the flat spacetime Minkowski vacuum, an accelerating detector will see a thermal spectrum but the regularized expectation value of stress-energy tensor will remain zero. For our purpose, it is clearly more meaningful to study $\langle T_{ab}\rangle_{\rm reg}$ rather than the detector response and we will concentrate on this study in this paper. However, in the last section of this paper, we will give the relevant results for the detector response, for the sake of comparison and completeness.   

The components of the stress-energy tensor depends on the choice of the coordinate system and hence could inherit the pathological characteristics of the coordinate system. It is therefore better to use physically well defined  \textit{scalar} quantities that have an invariant meaning. The most natural scalar quantities at any event in spacetime can be constructed in the freely falling frame at that event which eliminates any acceleration effects.  If we fill the spacetime by a suitable congruence of radially freely falling observers  with four-velocity $u^a$, then we can construct \cite{Ford1993, Ford1995, Ford1996, Visser1996a, Visser1996b, Visser1996c, Padmanabhan1987} two useful scalar quantities, at any event in spacetime, by the following definition:
\begin{equation}
\label{Paper3:Eq1}
\mathcal{U}=\langle \hat{T}_{ab}\rangle u^{a}u^{b};\qquad 
\mathcal{F}=-\langle \hat{T}_{ab}\rangle u^{a}n^{b}
\end{equation}
Here $n^a$ is the normal in the radial direction (such that $u_{a}n^{a}=0$), $\mathcal{U}$ is the energy density and $\mathcal{F}$ is the flux at that event as measured by a freely falling observer.  
      
As we said before, the most natural choice for $u^a$ is the four-velocity of the freely falling observers which are free of the acceleration effects. Further, we can fill the entire spacetime with the freely falling observers, which allows  a fairly uniform description of \emph{all} the regions of the spacetime. But, in principle, one can also define the scalars in Eq.~(\ref{Paper3:Eq1}) corresponding to \textit{any} observer with a given four-velocity $u^a$ and corresponding $n^a$. This is relevant in regions C and D of the spacetime where one can also introduce static observers and compute the energy density and flux as measured by them.

\begin{figure}[t!]
\centering
\includegraphics[scale=1.3]{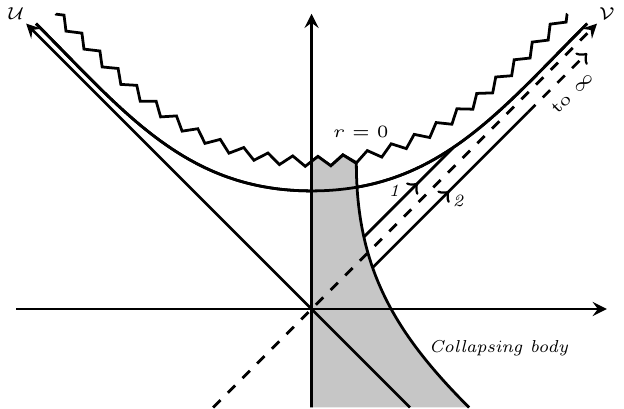}
\caption{The  collapse scenario in the Kruskal coordinates indicating  the $r=0$ and $r=\epsilon$ surfaces. Of the two null rays which are marked, the ray 1 reaches arbitrarily close to the singularity in finite time while the ray 2 propagates to asymptotic infinity at late times.
We will be interested in the energy density of the quantum field along the events in the path of ray 1, while ray 2 will be used for calibration. See text for detailed discussion.}
\label{Kruskal}
\end{figure}

Before proceeding further one needs to clarify couple of issues related to this approach. First, note that the $r=0$ singularity is a mathematically ill-defined event. Even though in the Kruskal coordinates in Fig.\ref{Kruskal} (both in (1+1) dimension and in (1+3) dimension) it is drawn as a hyperbola, with distinct $(T,X)$ coordinates along it, related by $T^{2}-X^{2}=1$ one needs to be careful while dealing with the mathematically ill-defined nature of this event. (For example, consider the two events on which the two null rays in bottom-left and bottom-right figures of Fig. \ref{Penrose} hit the singularity. One could worry whether these two events are physically distinct.) This is particularly true in the 
(1+3) dimension in which the angular part vanishes at $r=0$. 
To keep our discussion physically unambiguous, we will work throughout this paper with a surface infinitesimally close to $r=0$ but non-singular. Since $r\ne 0$ on this surface (see Fig. \ref{Kruskal}) any two separate points on it \emph{are indeed distinct}. Thus by working on a $r=\epsilon$ surface, with arbitrary small $\epsilon$ these issues  can be avoided. 

Second, the divergences in the \textit{densities} in the  $(1+3)$ dimension can sometimes be spurious because the volume factor can become arbitrarily small. So one can have a situation in which $\mathcal{U}$ diverges but $\mathcal{U}\sqrt{h}d^{3}x$ remains finite just because of geometrical considerations. So before we declare the energy density to be divergent in $(1+3)$
dimension we need to ensure that the total energy contained inside a small volume itself diverges, rather than just the energy density. This requires us to define a 3-volume element inside the event horizon. (One cannot, of course, use $4\pi r^2 dr$ because inside the event horizon, $r=\textrm{constant}$ surfaces are spacelike.) When we use a freely-falling observer with the four-velocity $u^a$ to define $\mathcal{U}$, the appropriate  3-volume measure to use is the one corresponding to the same observer in the synchronous coordinates. Using this measure and integrating $\mathcal{U}\sqrt{h}d^{3}x$  over a small volume around the singularity, we can  determine the proper nature of divergence near $r=0$. We will show later that the energy \textit{density} $\mathcal{U}$ diverges in both regions $A$ and $B$ in (1+1) spacetime dimensions. However in (1+3) dimension the divergence in region $A$ is compensated  by the shrinking volume measure (thereby making the energy inside a small region around the singularity 
finite), while the divergence in region $B$ persists, even though it is milder (being only $\epsilon ^{-4/3}$ compared to $\epsilon^{-2}$ in the (1+1) dimensions). 

We will work with the Oppenheimer-Snyder  model \cite{Oppenheimer1939, Vaz2007, Ross1993, Casadio1996, Poisson2004, Davies1976, Tolman1930}, which corresponds to  collapsing dust (matter with zero pressure) that forms a black hole. In this case, the spacetime inside the collapsing matter is described by the closed Friedmann metric. To compute the (regularised) stress-energy tensor of the field in a particular quantum state, we use the tools of the standard conformal field theory techniques. (We expect the results to map to the  s-wave sector of the scalar field in the collapsing background in $(1+3)$ case which is the usual assumption made in literature; see e.g., ref. \cite{Fabbri2005, Brout1995}) This essentially implies that at any point of the spacetime, we have ingoing and outgoing (spherical) waves and hence the whole manifold can be coordinatized using these null rays. The incoming ray comes directly from $\mathcal{J}^-$ and the outgoing ray comes from $\mathcal{J}^-$ after a reflection from the 
vertical line in the Penrose diagrams, representing $r=0$. As for the vacuum state, we shall work with the natural $in$-vacuum that is uniquely defined on $\mathcal{J}^-$. The relation of this state with the Unruh vacuum and its effects in the region C for the case of thin null shell collapse have been discussed in the refs. \cite{Suprit2014b, Suprit2014a, Smerlak2013} and will not be repeated here. 

\paragraph*{Summary of results.} Our study leads to the following results, particularly in the region \textit{inside} the the event horizon:

\begin{itemize}

\item The energy density and flux, as measured by the radially in-falling observers, both on the inside and outside the dust sphere (but inside the horizon) diverge on approaching the singularity. (This feature was also observed previously for a null shell collapse \cite{Suprit2014a}). Hence, the energy density for the scalar field can be arbitrarily large close to the singularity, even \textit{outside} the dust sphere. The same results hold on events along the null rays, inside the event horizon (shown in the bottom figures of \fig{Penrose}). For the ray reaching arbitrarily close to the singularity outside of the dust sphere (viz., the one shown in the bottom left of  \fig{Penrose}), both energy density and the flux diverge to arbitrarily high values as the ray approaches the singularity. Its implications remain to be ascertained but it could potentially affect the nature of the singularity (see e.g. \cite{Bardeen2014,Casadio2014}).

\item Further, the total energy obtained by integrating the energy density over a small three volume near the singularity also diverges in both (1+1) dimension [as $\epsilon ^{-2}$] and in the (1+3) dimension [as $\epsilon ^{-4/3}$]. This divergence  might prohibit the formation of singularity  due to arbitrarily large back-reaction from the quantum field. It should be noted that whenever we probe the null rays or in-falling observers that hit the singularity, we can always work with spacetime events \textit{arbitrarily close to} but distinct from the singularity, which have increasingly high curvature.

\item Let us consider the energy densities along the events in the rays shown in the top right and bottom left of  \fig{Penrose}. The spacetime events along these two rays, with one outside and one inside the event horizon, have approximately equal energy density when they are inside the matter.  However, as we probe the events along the inside ray,  which approaches singularity, the ratio of the energy densities on the outside events (which are approaching future null infinity) to the inside events (which are approaching the singularity) goes to zero due to arbitrarily high value of energy density, close to the singularity, on the latter null ray.

\item Similar behaviour is seen on the spacetime events along the null ray which are completely inside the dust sphere. In this case as well, as we approach  the singularity, we find a divergent energy density and flux.  The ratio $(\mathcal{U}/\rho)$ of the energy density of the scalar field, to that of the dust  also diverges as we approach the singularity in both the $(1+1)$ and $(1+3)$ spacetime dimensions. Hence the back-reaction due to the scalar field \emph{will} \emph{affect} the region inside the matter as well. 

\item As regards  the outside region, we find that the energy density and flux, measured by static observers and radially in-falling observers, exactly mimic the results obtained earlier in the literature in the  context of a null shell collapse. This suggests that these results are generic and independent of the nature of collapse. The energy density at the spacetime events along the null ray moving forward to future null infinity is always finite and reaches the standard value (corresponding to Hawking evaporation) at late times. This is in agreement with the earlier investigations.

\end{itemize}
The plan of the paper is as follows. We briefly review the matching of the interior and exterior parts of the spacetime for a collapsing dust ball in \sect{Paper3:SecGeom} and then introduce the double null coordinate spanning all of the spacetime. In \sect{Paper3:SecSET}, we consider the scalar field on this background and compute the components of the regularised stress-energy tensor for same. Subsequently, we compute the energy density and fluxes for various cases  in \sect{Paper3:SecENGFLUXTEMP}. Introducing the radially infalling observers, we study the behaviour of the invariant observables $\mathcal{U}$ and $\mathcal{F}$ along the three null rays mentioned earlier in \sect{Paper3:SecNULLALL}. Lastly, for the sake of completeness, we discuss the results regarding the detector response in terms of the effective temperature in \sect{effectemp}. The last section contains the concluding remarks.

\section{The Gravitational Collapse Geometry}\label{Paper3:SecGeom}

In this section, we briefly review the junction conditions for matching the interior Friedmann universe to the exterior Schwarzschild spacetime.We will then introduce a useful global coordinate syatem (which we call the double null coordinates) and express both the interior and exterior coordinates in terms of this double null coordinates in a conformally flat form, in the $(1+1)$ sector of the spacetime. This allows us to   determine the conformal factor of the metric from which we can calcualte  the regularised stress-energy tensor.

\subsection{Junction Conditions}\label{Paper3:SecJunc}

We consider the collapse of a spherical region filled with pressure-free dust in the (1+3) spacetime dimensions. The inside region of the dust sphere is homogeneous and isotropic and the metric interior to the dust will be a closed $(k=1)$ Friedmann model with the line element
\begin{equation}\label{Paper3:Eq2}
ds^{2}_{int}=-d\tau ^{2}+a^{2}(\tau)d\chi ^{2}+a^{2}(\tau)dL_{\perp}^{2}
\end{equation}
with $\tau$ denoting proper time of the dust particles, comoving with $x^{\mu}=\textrm{constant}$ and $dL_{\perp}^{2}=\sin ^{2}\chi d\Omega ^{2}$. A more convenient form for the line element can be obtained by transforming the proper time to conformal time $\eta$ via the relation:
\begin{equation}\label{Paper3:Eq3}
\eta =\int \frac{d\tau}{a(\tau)}
\end{equation}
in terms of which the interior metric reduces to the following form:
\begin{equation}\label{Paper3:Eq4}
ds^{2}_{int}=a^{2}(\eta)\left(-d\eta ^{2}+d\chi ^{2}+dL_{\perp}^{2}\right).
\end{equation}
The Einstein equation for the interior Friedmann universe filled with dust can be solved in a parametric form leading to the results:
\begin{align}
a(\eta)&=\frac{1}{2}a_{max}\left(1+\cos \eta \right)
\label{Paper3:Eq5a}
\\
\tau (\eta)&=\frac{1}{2}a_{max}\left(\eta +\sin \eta \right)
\label{Paper3:Eq5b}
\end{align}
In these conformal coordinates the surface of the dust sphere is taken to be located at some value $\chi =\chi _{0}$ and the collapse starts at $\eta =\tau =0$ and ends at $\eta =\pi ,\tau=(\pi/2)a_{max}$. The total energy contained within the dust sphere is constant and can be determined in terms of the quantity $a_{max}$ as:
\begin{equation}\label{Paper3:Eq6}
\rho a^{3}=\textrm{constant}=\frac{3}{8\pi}a_{max}
\end{equation}
The exterior region, which  is spherically symmetric and empty, is described by Schwarzschild metric but in a different set of coordinates. The spherical symmetry both inside and outside suggests that the angular coordinates for both the metric can be taken to be identical. Hence the outside line element has the following expression (we shall use the units with $2M=1$ henceforth):
\begin{equation}\label{Paper3:Eq7}
ds^{2}_{ext}=-\left(1-\frac{1}{r}\right)dt^{2}+\left(1-\frac{1}{r}\right)^{-1}dr^{2}+dL_{\perp}^{2}
\end{equation}
where we have $dL_{\perp}^{2}=r^{2}d\Omega ^{2}$. The above line element can also be written in terms of outgoing Eddington-Finkelstein coordinates $v=t+r^{*}$, with $r^{*}=r+\ln \left(r-1\right)$, such that the line element reduces to
\begin{equation}\label{Paper3:Eq8}
ds^{2}_{ext}=-\left(1-\frac{1}{r}\right)dv^{2}+2dvdr+dL_{\perp}^{2}
\end{equation}
In the outside coordinates the surface of the collapsing matter is characterized by $r=r(\tau)$ and $t=t(\tau)$, where, $r$ and $t$ are the Schwarzschild coordintes with $\tau$ being the internal time coordinate. The above surface is also determined by $\chi =\chi _{0}$. Therefore the connecting equation for radial coordinate is \cite{Padmanabhan2010b,Poisson2004}:
\begin{equation}\label{Paper3:Eq9}
r(\eta)=a(\eta)\sin \chi _{0}=\frac{1}{2}a_{max}\sin \chi _{0}\left(1+\cos \eta \right)
\end{equation}
Next we need to solve for the time coordinate, which is obtained by solving the following differential equation:
\begin{equation}\label{Paper3:Eq10}
\left(1-\frac{1}{r}\right)\frac{dt}{d\tau}=\textrm{constant}
\end{equation}
Matching the extrinsic curvature on both sides of the surface of the dust sphere we can fix the constant to be $\cos \chi _{0}$. This leads to the following differential equation for Schwarzschild time as:
\begin{equation}\label{Paper3:Eq11}
\frac{dt}{d\eta}=a_{max}\cos \chi _{0}\cos ^{2}\frac{\eta}{2}
+\frac{a_{max}\cos \chi _{0}\cos ^{2}\frac{\eta}{2}}{a_{max}\sin \chi _{0}\cos ^{2}\frac{\eta}{2}-1}
\end{equation}
Thus the time coordinate can be obtained from the above equation as:
\begin{align}\label{Paper3:Eq12}
t(\eta)&=\left\lbrace \left(1+\frac{a_{max}\sin \chi _{0}}{2}\right)\cot \chi _{0}\right\rbrace \eta
+a_{max}\cos \chi_{0}\sin \frac{\eta}{2}\cos \frac{\eta}{2}
\nonumber
\\
&+\ln \left \vert \frac{\sin ^{2}\frac{\eta}{2}-\cos ^{2}\frac{\eta}{2}+a_{max}\sin \chi _{0}
\cos ^{2}\frac{\eta}{2}+2\cot \chi _{0}\sin \frac{\eta}{2}\cos \frac{\eta}{2}}
{\left(a_{max}\sin \chi _{0}\cos ^{2}\frac{\eta}{2}-1\right)} \right \vert
\end{align}
The time corresponding to horizon crossing of the collapsing surface can be obtained by setting $r=1$, which leads to, $\eta _{H}=\pi -2\chi _{0}$. Note that as $\eta \rightarrow \eta _{H}$, $t(\eta)$ diverges; thus, for an outside observer, the surface of the imploding matter takes infinite time to reach the event horizon. The $a_{max}$ and $\chi _{0}$ are not independent, since mass of the imploding dust ball is:
\begin{equation}\label{Paper3:Eq13}
M=\frac{4\pi}{3} \rho r^{3}=\frac{4\pi}{3}\rho a^{3}\sin ^{3}\chi _{0}
=\frac{1}{2}a_{max}\sin ^{3}\chi _{0}
\end{equation}
Thus with $2M=1$ we are left with the condition $a_{max}\sin ^{3}\chi _{0}=1$. With these conditions we find the connection between the Eddington-Finkelstein coordinate $v$ in the exterior region to that of interior region to be:
\begin{align}
v(\eta)&=\left\lbrace \cot \chi _{0}\left(1+\frac{a_{max}\sin \chi _{0}}{2}\right)  \right\rbrace \eta
+a_{max}\sin \chi _{0}\left[\cos \frac{\eta}{2}+\cot \chi _{0}\sin \frac{\eta}{2} \right]
\cos \frac{\eta}{2}
\nonumber
\\
&+2\ln \left \vert \sin \frac{\eta}{2}+\cot \chi _{0}\cos \frac{\eta}{2} \right \vert
\label{Paper3:Eq14}
\end{align}
Thus we have the matching condition of the outside Schwarzschild coordinates with the inside Friedmann  coordinates. With this matching condition we now proceed to determine the double null coordinates for the entire spacetime region.

\subsection{Double Null Coordinates}\label{Paper3:SecNull}

\begin{figure}[t!]
\centering
\includegraphics[scale=1]{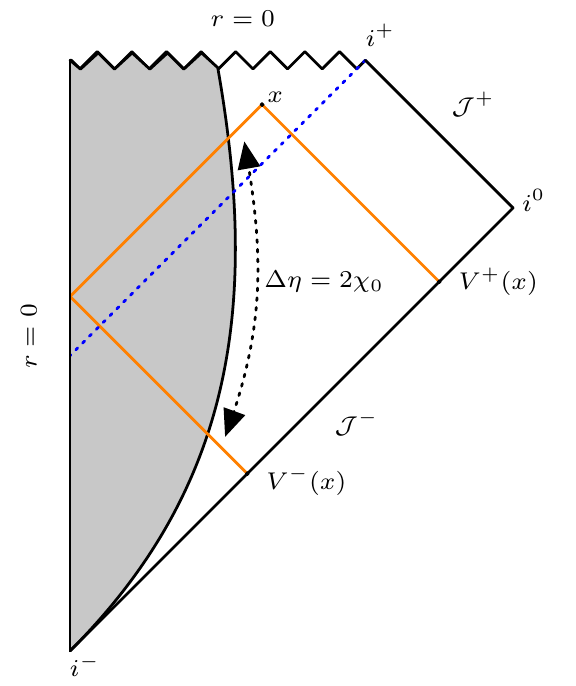}
\caption{Penrose diagram showing the construction of double null coordinates. Each event $x$ can be characterized by two null rays --- an incoming ray from $\mathcal{J}^{-}$, which is labeled as $V^{+}$ and an outgoing ray, tracked backward to the vertical line $r =0$ and then reflected towards $\mathcal{J}^{-}$, giving the second null coordinate $V^{-}$.}
\label{coordinates}
\end{figure}

In this and subsequent sections we will first concentrate on the (1+1) spacetime dimensions by using $dL_{\perp}^{2}=0$ (which corresponds to $d\theta =0$ and $d\phi =0$) and quote the results for 1+3 dimensions, whenever appropriate. For the purpose of calculating the vacuum expectation values for the regularised stress-energy tensor, it is useful to  define and use a coordinate system made up of double null coordinates, constructed from the ingoing and outgoing null rays. Each event $x$ (see \fig{coordinates}) can be characterized by two null rays --- an incoming ray from $\mathcal{J}^{-}$, which is labeled as $V^{+}$ and an outgoing ray, tracked backward to vertical line $r =0$ and then reflected to $\mathcal{J}^{-}$ giving the second null coordinate $V^{-}$. To obtain these null coordinates for the global spacetime we proceed as follows: In the interior Friedmann region we define two new coordinates which are both null
\begin{equation}
\label{Paper3:Eq15}
U=\eta -\chi ;\qquad V=\eta +\chi
\end{equation}
Then the line element with the coordinates $(V,U)$ in the interior Friedmann universe reduces to the following form:
\begin{equation}\label{Paper3:Eq16}
ds^{2}_{int}=-a^{2}(\eta)dUdV
\end{equation}
In order to describe the exterior structure we define ingoing and outgoing Eddington-Finkelstein coordinates $v$ and $u$ both being null. The exterior line element  reduces to:
\begin{equation}\label{Paper3:Eq17}
ds^{2}_{ext}=-\left(1-\frac{1}{r}\right)dudv
\end{equation}
in the null coordinates $(v,u)$.
Then we can define the double null coordinates $V^{+}$ and $V^{-}$ through the following relations related to the interior Friedmann universe as:
\begin{align}
V^{+}=A(V-\chi _{0})
\label{Paper3:Eq18a}\\
V^{-}=A(U-\chi _{0})
\label{Paper3:Eq18b}
\end{align}
where we have introduced the function $A(x)$ for future convenience:
\begin{align}
A(x)&\equiv \left\lbrace \cot \chi _{0}
\left(1+\frac{a_{max}\sin \chi _{0}}{2}\right)  \right\rbrace 
x+
2\ln \left \vert \sin \frac{x}{2}+ \cot \chi _{0}\cos \frac{x}{2} \right \vert
\nonumber
\\
&+a_{max}\sin \chi _{0}\left[\cos \frac{x}{2}
+\cot \chi _{0}\sin \frac{x}{2} \right]
\cos \frac{x}{2}
\end{align}
which connect the interior coordinates $(U,V)$ to the global double null coordinates $(V^{+},V^{-})$. 

From the construction it is easy to verify the following results. 
The surface $r=0$, which is in the FRW universe, has equation $\chi =0$ or equivalently (from \eq{Paper3:Eq15}) $U=V$. Thus from \eqs{Paper3:Eq18a} and (\ref{Paper3:Eq18b}) it is evident that $V^{+}=V^{-}$. Thus the two null coordinates coincide at $r=0$. Also these null coordinates should be continuous on the surface. Then using the fact that $V=\eta +\chi _{0}$ on the surface of the star from \eq{Paper3:Eq18a} we get, $V^{+}=v$, which it should. For  $\chi \neq 0$, the value of $V^{+}$ and $V^{-}$ are never equal for all $\eta$. This implies that even though the singularity forms at $\eta =\pi$, in the double null coordinates it is `spread out'. The value of $V^{-}$ for the outer surface of the star to reach the singularity is obtained by using $U=\pi -\chi _{0}$, leading to, $V^{-}=A(\pi-2\chi _{0})$. By construction this must be equal to $V^{+}$ of the point, where the ray enters the dust sphere, which is $A(\eta)$. This helps us to identify, the time at which that particular null ray has entered the 
sphere as $\pi -2\chi _{0}$. This, in turn, fixes the surface of the star completely. Note that this is just a corollary of a more general result: The null rays entering the sphere, getting reflected at $r=0$ and exiting the sphere satisfy the relation, $d\eta =d\chi$. Thus the ray starts at  $\chi =\chi _{0}$, then goes to $\chi =0$, again comes out at $\chi =\chi _{0}$. The difference in the coordinate values for $\eta$ between the point of entering and exit of the ray corresponds to $\Delta \eta =2\chi _{0}$ (see Fig. \ref{coordinates}). Hence in the above situation, the final value of $\eta$ is $\pi$, and the entry value of $\eta$ has to be $\pi -2\chi _{0}$, leading back to the previous result. 
  
Another set of relations connecting exterior coordinates $(u.v)$ to the global null coordinates $(V^{+},V^{-})$ are given by:
\begin{align}
v&=V^{+}
\label{Paper3:Eq19a}
\\
u&=B(U+\chi _{0})=A(U+\chi _{0})-a_{max}\sin \chi _{0}\left(1+\cos (U+\chi _{0}) \right)
\nonumber
\\
&-2\ln \left \vert a_{max}\sin \chi _{0}\cos ^{2}\left(\frac{U+\chi _{0}}{2} \right)-1 \right \vert
\label{Paper3:Eq19b}
\end{align}
Thus the interior and exterior line elements can be written in the double null coordinate system $(V^{+},V^{-})$ in the following manner:
\begin{align}
ds^{2}_{int}&=-a^{2}\left(\frac{U+V}{2}\right)\frac{1}{\frac{dA}{dU}\frac{dA}{dV}}dV^{+}dV^{-}
\label{Paper3:Eq20a}
\\
ds^{2}_{ext}&=-\left(1-\frac{1}{r}\right)\frac{dB/dU}{dA/dU}dV^{+}dV^{-}
\label{Paper3:Eq20b}
\end{align}
Hence this double null coordinate covers the full spacetime and brings the $(1+1)$ sector of the spacetime to the conformally flat form. This is especially suited to evaluate the vacuum expectation value of the regularised stress-energy tensor, which is our next task.

\section{Regularised Stress-Energy Tensor}\label{Paper3:SecSET}

We consider a minimally coupled, massless scalar field on the background geometry described by the line elements  in \eqs{Paper3:Eq20a} and (\ref{Paper3:Eq20b}). In two dimension the dynamics of the geometry is encoded in the conformal factor which allows us to obtain the vacuum expectation value of the energy momentum tensor for the scalar field. For this we follow the standard procedure \cite{Davies1976,Brout1995} and use the following expressions given in terms of the conformal factor:
\begin{align}
\langle T_{++}\rangle &=\frac{1}{12\pi}\left[\frac{1}{2}\frac{\partial _{+}^{2}C}{C}
-\frac{3}{4}\left(\frac{\partial _{+}C}{C}\right)^{2} \right]
\label{Paper3:Eq22a}
\\
\langle T_{--}\rangle &=\frac{1}{12\pi}\left[\frac{1}{2}\frac{\partial _{-}^{2}C}{C}
-\frac{3}{4}\left(\frac{\partial _{-}C}{C}\right)^{2} \right]
\label{Paper3:Eq22b}
\\
\langle T_{+-}\rangle &= \frac{1}{24\pi}\left[\frac{\partial _{+}\partial _{-}C}{C}
-\frac{\partial _{+}C}{C}\frac{\partial _{-}C}{C}\right]
\label{Paper3:Eq22c}
\end{align}
In the above expressions we have introduced a short hand notation: the symbol 
$\pm$ stands for $V^{\pm}$ respectively. The detailed computation and the
explicit expression for the regularised stress-energy tensor, using these relations is given
in the Appendix \ref{Paper3:AppEMT}.

\section{Energy Density and Flux observed by Different Observers}\label{Paper3:SecENGFLUXTEMP}

We shall now compute the energy density and flux at different events in the spacetime using the regularised stress tensor. As we described earlier our primary interest is in the regions B and A inside the event horizon where no static observers can exist. In these regions,  we will study the energy density and the flux in the freely-falling frame. But before we do that, it is useful to consider the region C and see how the standard results are reproduced. In this region C (unlike in A and B) we can introduce both static and freely falling observers and study the flux and energy density as measured by both kinds of observers. 

\subsection{Static Observers in region C}

We will start with the energy density and flux as measured by the static observers in region C. This will verify our procedure by leading to the standard result of the Hawking radiation with the Tolman redshifted temperature. 

\begin{figure*}
\begin{center}

\includegraphics[height=2in, width=3in]{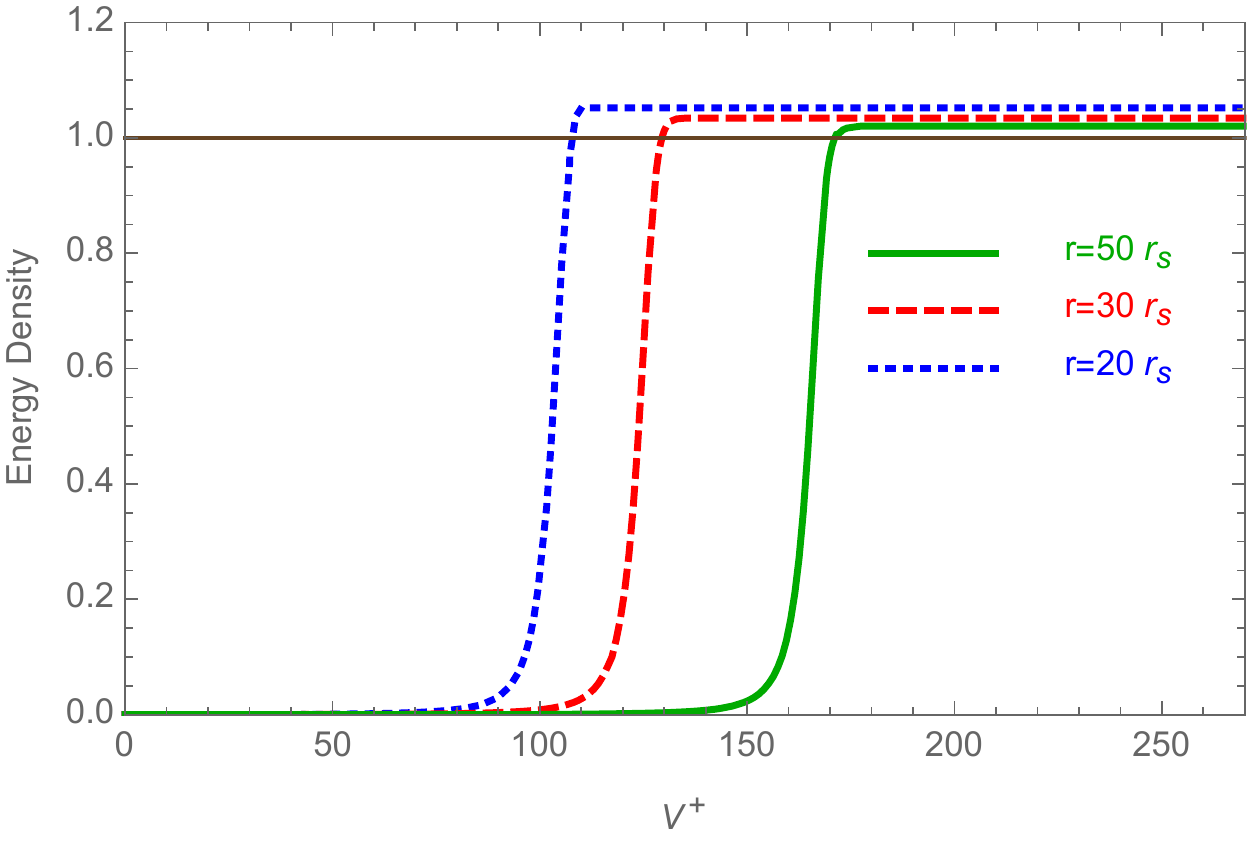}~~
\includegraphics[height=2in, width=3in]{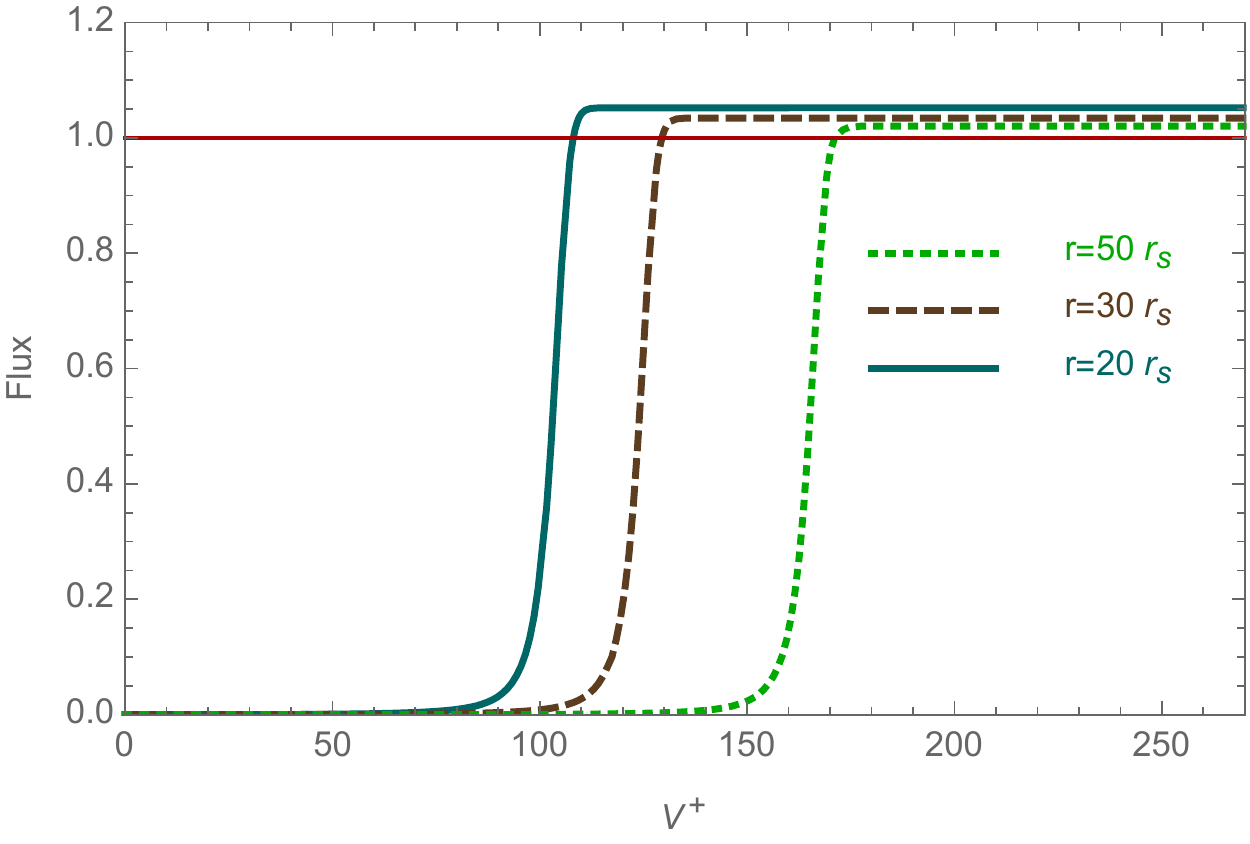}\\

\caption{Variation of the  energy density and the flux  as observed 
by a static observer as a function of the proper time. The first figure shows the variation of the 
energy density  with proper time $V^{+}$. 
All the static observers at different radii will ultimately observe 
the standard result for the Hawking radiation with the temperature modified by the Tolman redshift factor. (Note that while 
plotting the graphs we have normalized them to their values at infinity, i.e. 
$\pi T_{H}^{2}/12$ for convenience.) The same situation is depicted in the second figure 
for the flux. Both  start with a  small value and then rise 
rapidly, ultimately saturating at the appropriate values for the  Hawking radiation with the redshifted temperature.
}\label{Paper3:FigSTAT01}

\end{center}
\end{figure*}

A static observer  stays outside the event horizon 
at some fixed radius $r>1$. Since the observer is not following 
geodesic he/she has to accelerate (by firing rockets) so as to remain stationary 
at that radius. The velocity components for this static observer is 
given by:
\begin{align}
\dot{V}^{+}&=\sqrt{\frac{r}{r-1}}
\label{Paper3:Eq26a}
\\
\dot{V}^{-}&=\sqrt{\frac{r}{r-1}}
\left(\frac{\cot \chi _{0}-\tan \left(\frac{U+\chi _{0}}{2}\right)}
{\cot \chi _{0}+\tan \left(\frac{U-\chi _{0}}{2}\right)}\right)
\frac{\cos ^{2}\left(\frac{U-\chi _{0}}{2}\right)}{\cos ^{2}\left(\frac{U+\chi _{0}}{2}\right)}
\label{Paper3:Eq26b}
\end{align}
for a fixed $r$. The outward normal, determined from the condition 
$n_{a}\dot{V}^{a}=0$, has the following components:
\begin{equation}\label{Paper3:Eq27}
n^{+}=\dot{V}^{+};\qquad n^{-}=-\dot{V}^{-}
\end{equation}
Thus the energy density and flux can be computed using \eq{Paper3:Eq1} for the static 
trajectories as a function of $V^{+}$. For static observers $V^{+}$ acts as the proper time 
along their trajectories and we plot both the energy density and flux as a function of $V^{+}$ in \fig{Paper3:FigSTAT01}. From this figure we see that both energy density and flux shows similar 
behaviour at large radii. Both of them start from  small positive values and finally 
saturate at the  standard thermal spectrum values with the Tolman redshifted value for the temperature. 

The key results we want to verify for the static observers are, of course, the late time energy 
density and flux. The late time limit corresponds to $U\rightarrow \pi -3\chi _{0}$, 
under which both the energy density and the flux lead to the following expressions:
\begin{subequations}
\begin{align}
\mathcal{U}&=\frac{\pi T_{H}^{2}}{12}\left(1-\frac{2}{r^{4}}\right)
\left(\frac{r}{r-1}\right)
\label{Paper3:Eq28a}
\\
\mathcal{F}&=\frac{\pi T_{H}^{2}}{12}\frac{r}{r-1}
\label{Paper3:Eq28b}
\end{align}
\end{subequations}
We see  that, in the asymptotic limit, $r\rightarrow \infty$ (corresponding to a static observer at a large distance) above expressions reduce to the standard Hawking energy density and flux. The redshift  factor, as is well-known, 
diverges as the horizon $r=1$ is approached. From \eq{Paper3:Eq28a} it is evident 
that in the near horizon limit the energy density reaches a maximum and then decreases 
with decreasing $r$, eventually becoming negative. Thus, even in the Oppenheimer-Snyder dust model there is a region of negative energy density just outside the black hole horizon, just as in the case of a null shell collapse noticed earlier e.g., in Ref. \cite{Suprit2014a}. However the flux  is always positive and diverges at the horizon. 

\subsection{Radially In-falling Observers}

Having reproduced the standard results in region C we now turn our attention to the phenomena inside the event horizon. Since there can be no static observers in this region, it is best to study these effects in the freely falling frame of the radially ingoing observer. To provide the complete picture and to maintain continuity, we will study the results for the freely falling observers in the \textit{entire} spacetime.
This can be done from two different perspectives:

\begin{itemize}
\item[(1)] We can compute the energy density and flux on the events  along  the trajectory of a given freely falling observer. This has a clear physical meaning of what such a freely falling observer will see, as she plunges into the singularity.

\item[(2)] We can examine the energy density and flux on the events along specific null rays we described in \sect{Sec01n} (see Fig. \ref{Penrose}). This will tell us how the energy density and flux behaves on a selected set of events in the spacetime which are physically well-motivated.
\end{itemize}
In the section, we shall present the results based on the approach (1), viz. along the trajectories of the observers. We shall describe the behaviour of the energy density and flux on the events along the specific null rays in \sect{Paper3:SecNULLALL}.

Since part of the spacetime is covered by the collapsing dust ball, there exist two types of radially in-falling 
observers. The first set of  observers are (i) those who are comoving with the dust sphere and remain \textit{inside} the collapsing matter, while the second set of observers are (ii) those who are always
 \textit{outside} the dust sphere. The first set of  observers who stay inside the 
dust sphere ultimately reach the singularity in a finite proper time. These observers 
are in a Friedmann universe. The observers in a region outside the collapsing matter also hit the singularity in finite time but they live in a 
Schwarzschild geometry. Thus these two observers are geometrically distinct even though 
both of them hit the singularity in a finite proper time. We will now consider these two sets of freely falling observers and use them to probe the regions D, A and B.

\subsubsection{Radially In-falling Observers: Inside regions D and A}
\label{RadFallIn}
\begin{figure*}
\begin{center}

\includegraphics[height=2in, width=2.8in]{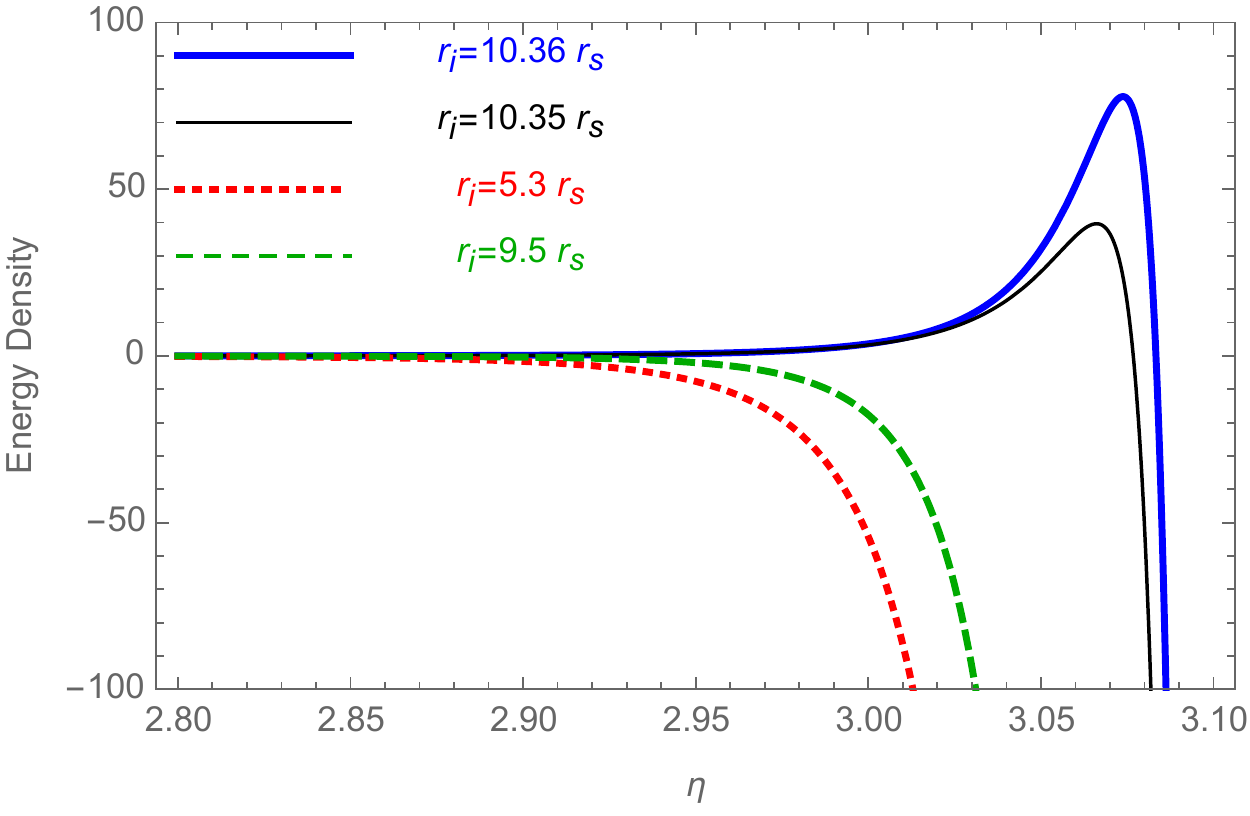}

\caption{The  variation of the energy density as the collapse progresses for an observer inside the sphere 
and co-moving with the sphere. The energy density diverges as the conformal time 
$\eta$ tends to $\pi$, the instant when the dust sphere collapses to a singularity.}
\label{Paper3:FigInfallENG01}

\end{center}
\end{figure*}

Let us start with the observers who are inside the dust sphere and are comoving with it.  
All these observers start at $\eta =0$ when the dust sphere starts to collapse and 
reach $\eta =\pi$ in a finite proper time. The trajectory of these observers are 
characterized by the conformal time $\eta$. For such an observer comoving 
with the dust sphere, the four velocity is given by:
\begin{align}
u^{a}&=\frac{1}{\sqrt{2}a\left(\eta \right)}\left(\frac{dA}{dV},\frac{dA}{dU}\right)
\label{Paper3:Eq31a}
\\
u_{a}&=-\frac{a}{\sqrt{2}}\left(\frac{1}{dA/dV},\frac{1}{dA/dU}\right)
\label{Paper3:Eq31b}
\end{align}
and the unit normal has the following expression:
\begin{align}
n_{a}&=\frac{a}{\sqrt{2}}\left(\frac{1}{\frac{dA}{dV}},-\frac{1}{\frac{dA}{dU}}\right)
\\
n^{a}&=\frac{1}{\sqrt{2}a}\left(\frac{dA}{dV},-\frac{dA}{dU}\right)
\end{align}
The comoving observer is characterized by the value of $\chi$ which remains fixed 
throughout the trajectory (taken as $\tilde{\chi}$) and 
hits the singularity as $\eta$ varies from $0$ to $\pi$. The variation of energy density 
for the comoving observer with the proper time $\eta$ along the trajectory 
is shown in \fig{Paper3:FigInfallENG01}. We note that the energy 
density for the outermost observer at $\chi =\chi _{0}$ remains positive 
and diverges as $\eta \rightarrow \pi$. While the energy density as 
measured by other observers remain finite while reaching a maximum near 
$\eta =\pi$ and then becomes negative. Finally the energy densities for 
all the observers diverge in the $\eta \rightarrow \pi$ limit. This result is shown in \fig{Paper3:FigInfallENG01}.

The energy density diverges as the observer hits the singularity. The 
ratio of the energy density measured by the observer to that of the 
energy density of the collapsing matter also diverges. This divergence can be obtained from the leading order behaviour of $\mathcal{U}$ and $\rho$ near $r=0$, which can be represented, for (1+1) spacetime dimensions, as: (see \eq{Coveq02} in the Appendix \ref{Paper3:AppENGFluxRadialIn})
\begin{equation}\label{Newadd01}
\mathcal{U}^{(1)}=-\frac{\kappa ^{2}a_{max}}{6\pi a^{3}};\qquad \rho ^{(1)} =\frac{a_{max}\sin ^{2}\chi _{0}}{2a}; \qquad \frac{\mathcal{U}^{(1)}}{\rho ^{(1)}}\propto \frac{1}{a^{2}}
\end{equation}
where superscript $(1)$ denote the values of respective quantities in (1+1) spacetime dimensions. 
Thus, close to the singularity, the energy density of the scalar field dominates over that of the dust sphere. 

We next consider the total energy within a small volume with linear dimension $\epsilon$ in this (1+1) spacetime both for the scalar field ($\mathcal{E}^{(1)}$) and classical matter ($E^{(1)}$). This  energy is given by:
\begin{align}
\mathcal{E}^{(1)}=\int \mathcal{U}\sqrt{h^{(1)}}dV^{(1)}\sim \frac{1}{\epsilon};\qquad
E^{(1)}=\int \rho \sqrt{h^{(1)}}dV^{(1)}=\epsilon;\qquad \frac{\mathcal{E}^{(1)}}{E ^{(1)}}=\frac{1}{\epsilon ^{2}}
\end{align}
Thus even the total energy of the scalar field within a small volume diverges and is much larger compared to the total energy of the classical matter within that small volume.  Energy in the quantum field dominates over the energy  of the classical background. 

\paragraph*{Generalization to (1+3) spacetime dimensions}

Let us now consider the generalization to (1+3) spacetime dimensions. Since we are considering s-wave approximation, the energy density for the scalar field in (1+1) spacetime dimensions can be related to that in (1+3) spacetime dimensions through the result \cite{Fabbri2005,Brout1995}: $\mathcal{U}^{(1)}\times(1/4\pi r^{2})=\mathcal{U}^{(3)}$, where $\mathcal{U}^{(1)}$ and $\mathcal{U}^{(3)}$ are the energy densities in (1+1) and (1+3) spacetime dimensions respectively.
So, inside the dust sphere, we have the following expressions for the energy densities of the scalar field and the dust:
\begin{align}
\mathcal{U}^{(3)}=-\frac{\kappa ^{2}a_{max}}{24\pi ^{2} a^{5}\sin ^{2}\chi _{0}};\qquad \rho ^{(3)} =\frac{3}{8\pi}\frac{a_{max}}{a^{3}}; \qquad \frac{\mathcal{U}^{(3)}}{\rho ^{(3)}}\propto \frac{1}{a^{2}}
\end{align} 
where the superscript $(3)$ denotes the values of the respective quantities in (1+3) spacetime dimensions.
Hence the divergence in $\mathcal{U}/\rho$ still persists in (1+3) spacetime dimensions.

As we have argued before, the really important measure is probably not the energy density but the total energy contained in a small volume of size $\epsilon$ around the singularity in the $\epsilon\rightarrow 0$ limit.  From \eq{Paper3:Eq2} we arrive at the volume element to be: $\sqrt{h}d^{3}x=a^{3}\sin ^{2}\chi \sin \theta ~d\chi d\theta d\phi$. The total  energy inside a small volume can be found by integrating the  energy density inside a sphere and   is given by (with $\epsilon \rightarrow 0$):
\begin{align}
\mathcal{E}^{(3)}=\int _{0}^{\epsilon} \mathcal{U}^{(3)}\sqrt{h^{(3)}}d^{3}x\sim \epsilon ;\qquad
E^{(3)}=\int _{0}^{\epsilon} \mathcal{\rho}^{(3)}\sqrt{h^{(3)}}d^{3}x \sim \epsilon ^{3};\qquad
\frac{\mathcal{E}^{(3)}}{E^{3}}=\frac{1}{\epsilon ^{2}}
\label{eratio}
\end{align}
It vanishes for both components (as $\epsilon ^{3}$ for the dust and as $\epsilon$ for the field), where $\epsilon$ represents radius of the small sphere around the singularity.
However the total energy in the scalar field dominates over  that in the classical background. This suggests  that inside the dust sphere, the effect of back-reaction can \emph{not} be neglected  in (1+3) spacetime dimensions. (In normal units, the ratio of the energy densities in 
\eq{eratio} will go as $L_P^2/\epsilon^2$; so the effect is significant numerically at Planck scales, which is not unexpected.)
 
\subsubsection{Radially In-falling Observers: Outside regions C and B}
\label{RadFallOut}
\begin{figure*}
\begin{center}

\includegraphics[height=2in, width=3in]{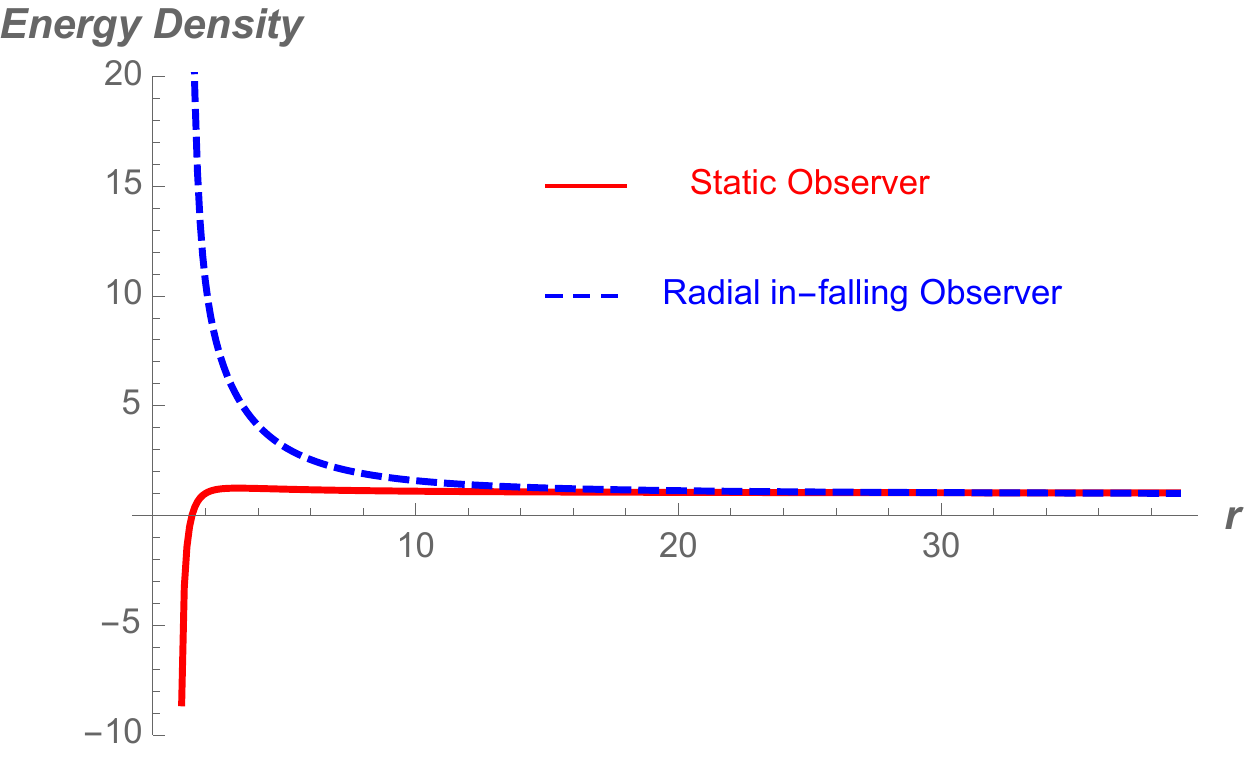}~~~
\includegraphics[height=2in, width=3in]{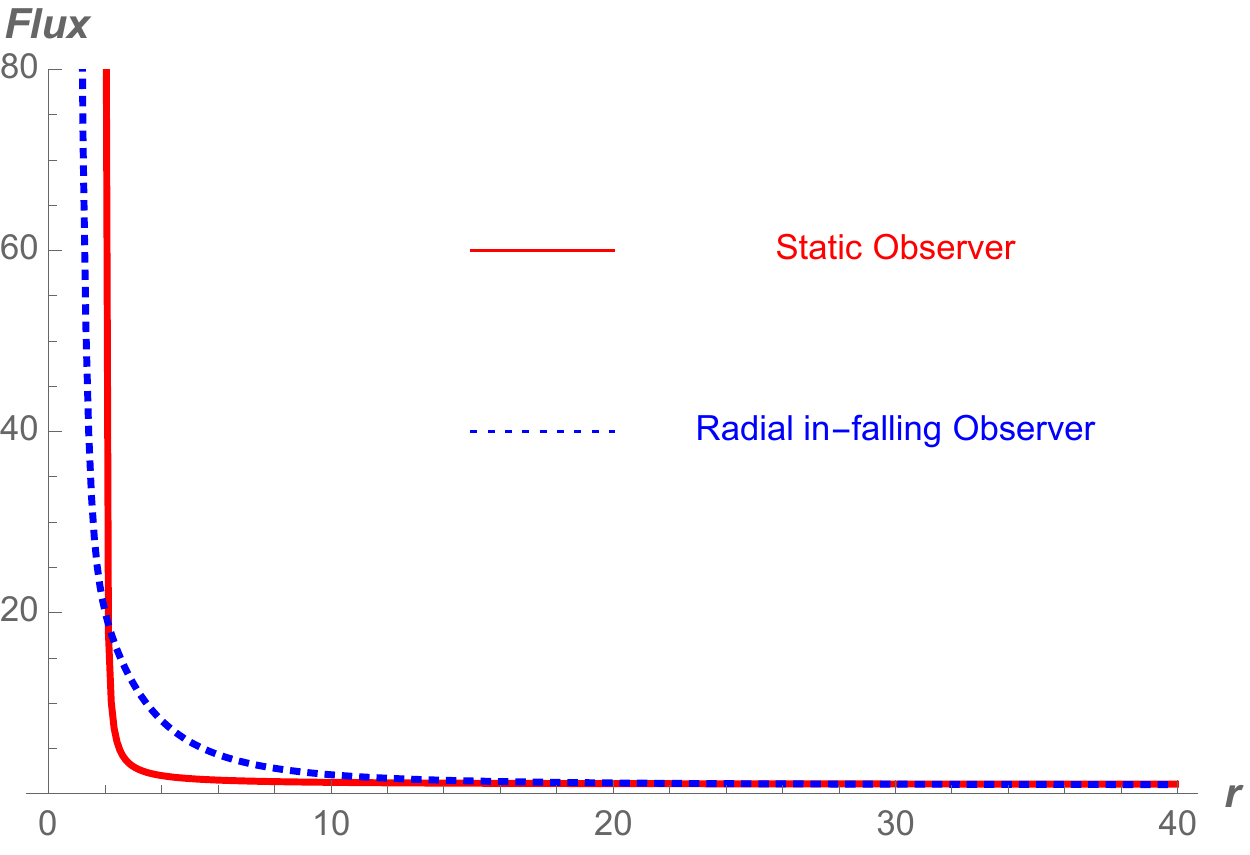}

\caption{The first figure shows the energy density for the static observers at various radii at late times 
compared to the radially in-falling observer who crosses those static observers at different times.
The second figure depicts the corresponding results for the flux. For static observers near 
the horizon the flux at late time diverges to infinity as the horizon is approached, 
while the energy density becomes negative. For radially in-falling observers nothing peculiar happens at $r=2M$, while both $\mathcal{U}$ and $F$ diverge as $r\rightarrow 0$. This feature was also observed earlier for the collapse of a null shell in \cite{Suprit2014a}.}
\label{Paper3:FigRadial01}

\end{center}
\end{figure*}

The last set of observers we will study are the radially in-falling ones, but outside the 
dust sphere. They start from some initial radius and then follow a geodesic, 
ultimately hitting the singularity. They, however, live in the Schwarzschild 
region of spacetime throughout their life. Any such in-falling observer is parametrised 
by the energy associated with the geodesic, or --- equivalently --- by the initial radius from 
which her suicide mission starts. These two quantities are related via:
\begin{equation}\label{Paper3:Eq33}
E=\left(1-\frac{1}{r_{i}}\right)^{1/2}
\end{equation}
where $r_{i}$ stands for the initial radius from which the free-fall 
begins. From the geodesic equation we can determine the four velocity 
of these observers, which turns out to be:
\begin{align}
\dot{V}^{+}&=\frac{r}{r-1}\left(E-\sqrt{E^{2}-\frac{r-1}{r}}\right)
\label{Paper3:Eq34a}
\\
\dot{V}^{-}&=\frac{r}{r-1}\left(E+\sqrt{E^{2}-\frac{r-1}{r}}\right)
\left(\frac{\cot \chi _{0}-\tan \left(\frac{U+\chi _{0}}{2}\right)}
{\cot \chi _{0}+\tan \left(\frac{U-\chi _{0}}{2}\right)}\right)
\frac{\cos ^{2}\left(\frac{U-\chi _{0}}{2}\right)}{\cos ^{2}\left(\frac{U+\chi _{0}}{2}\right)}
\label{Paper3:Eq34b}
\end{align}
The components of the normal are determined by the condition 
$n_{a}V^{a}=0$ which leads to the following choice for the 
outward normal:
\begin{equation}\label{Paper3:Eq35}
n^{+}=\dot{V}^{+};\qquad n^{-}=-\dot{V}^{-}
\end{equation}
However we also require the evolution of $V^{+}$ as the observer proceeds towards the 
singularity, i.e. we need $V^{+}$ as a function of the observer's conformal time $\eta$. 
This can also be determined from  the geodesic equation. 
For the radially in-falling trajectory the solution to the geodesic equation can be 
written as:
\begin{subequations}
\begin{align}
r(\eta)&=\frac{r_{i}}{2}\left(1+\cos \eta \right)
\label{Paper3:Eq36a}
\\
\tau &=\tau _{i}+\sqrt{r_{i}}\frac{r_{i}}{2}\left(\eta +\sin \eta \right)
\label{Paper3:Eq36b}
\end{align}
\end{subequations}
In this case as well, the conformal time $\eta$ varies in the same range $0\leq \eta \leq \pi$. 
We also have the following initial conditions: $r(\tau _{i})=r_{i}$. 
From these two relations we can determine $d\tau /d\eta$ and, on using \eq{Paper3:Eq34a}, we get the differential equation satisfied by $V^{+}(\eta)$:
\begin{equation}\label{Paper3:Eq37}
\frac{dV^{+}}{d\eta}=r_{i}\cos \left(\frac{\eta}{2}\right) 
\frac{\sqrt{r_{i}}\left[1-\left(1/r_{i}\right)\right]^{1/2}
\cos \left(\eta /2\right)-\sin \left(\eta /2\right)}
{1-\left\lbrace \sec ^{2}\left(\eta /2\right)/r_{i}\right\rbrace} 
\end{equation}
The above equation can be integrated to obtain $V^{+}$ as a function of 
the time $\eta$. Then both the energy density and flux can be obtained 
from this prescription. The behaviour of both the energy density 
and the flux are shown in \fig{Paper3:FigRadial01}. 
We 
note that both the energy density and the flux show identical behaviour. 
For $\eta <0$, when the sphere has not started collapsing, the energy 
density and flux vanish. (With our choice $2M=1$ the asymptotic value of Hawking flux would turn out to be: $(1/192\pi)$. To normalize the figures so that Hawking flux turns out be unity, the figures are drawn with a rescaling of the y-axis.) As the observer moves forward in  
proper time both the energy density and the flux arise and  
--- as the event horizon is approached --- both the energy density and the flux rise rapidly. However the values are finite at the event horizon located at 
$\eta _{H}\sim 2.824$. This suggests another interesting aspect to study, which is the comparison between the energy density and flux as measured by a radially in-falling observer and various static observers they encounter along the path. For the static observers the energy density and flux 
diverge as the event horizon is approached, while for the radially in-falling observer 
the energy density and flux remain finite at the horizon crossing. 
We give below the expressions for the energy density and flux as measured by these observers 
at the horizon: 
\begin{subequations}
\begin{align}
\mathcal{U}_{H}&=\pi T_{H}^{2}\left(\frac{2}{3}-\frac{1}{48 E^{2}}+2E^{2}\right)
\label{Paper3:Eq38a}
\\
\mathcal{F}_{H}&=\pi T_{H}^{2}\left(\frac{1}{48 E^{2}}+2E^{2}\right)
\label{Paper3:Eq38b}
\end{align}
\end{subequations}
For a radially in-falling observer from infinity we have $E=1$, for which the horizon crossing energy density and flux becomes $\mathcal{U}_{H}\sim 32 \mathcal{U}_{\infty}$ and $\mathcal{F}_{H}\sim 24 \mathcal{F}_{\infty}$. (Note that these results were obtained earlier in the context of a null collapse in \cite{Suprit2014a}.)

However the energy density and flux finally diverge as the singularity is approached. The behaviour of the energy density and its integral over a small volume has the following expressions near the singularity: (see \eq{EngAppTotal} in Appendix \ref{Paper3:AppENGFluxRadialOut})
\begin{align}
\mathcal{U}^{(1)}=\frac{7\kappa ^{2}}{24\pi}\frac{1}{\epsilon^{3}};\qquad
\mathcal{E}^{(1)}=\int \mathcal{U}\sqrt{h^{(1)}}dV^{(1)}\sim \frac{1}{\epsilon^{2}}
\end{align}
where superscript $(1)$ denotes the expression of the quantities in the (1+1) spacetime dimensions. Thus both the energy density and the integrated energy diverges as the singularity is approached. Hence even in region \emph{outside} the dust sphere the backreaction is important near the singularity.

\paragraph*{Generalization to (1+3) spacetime dimensions} The energy density of the scalar field, as measured by the radially in-falling observers, in (1+3) spacetime dimensions can be obtained by dividing the (1+1) energy density by $4\pi r^{2}$. (This is a standard procedure adopted in the literature; see e.g., \cite{Fabbri2005,Brout1995}). This leads to the following approximate expression for the energy density near the singularity:
\begin{align}
\mathcal{U}^{(3)}=\frac{7\kappa ^{2}}{96\pi ^{2}}\frac{1}{r^{5}}
\end{align}
This result, at the face of it, shows that the back-reaction effects will be quite significant close to the singularity even \emph{outside} the collapsing dust sphere. However, we need to ensure that the geometrical factor arising from the shrinking of the spatial volume does not over-compensate the divergence. To obtain the proper volume element appropriate for the radially in-falling observer, let us start with the  metric in the synchronous coordinates for an observer in free-fall, from a large distance:
\begin{align}
ds^{2}=-d\tau ^{2}+\frac{1}{r}dR^{2}+r^{2}d\Omega ^{2};\qquad r=\left[\frac{3}{2}\left(R-\tau \right)\right] ^{2/3}
\end{align}
Thus for a $\tau =\textrm{constant}$ surface the volume element turns out to be:
\begin{align}
\sqrt{h}d^{3}x=r^{3/2}\sin \theta ~dR d\theta d\phi =\frac{3}{2}\left(R-\tau \right)\sin \theta dR d\theta d\phi 
\end{align}
Integrating this energy density over a sphere of small radius $\epsilon$ we get the total energy to be 
\begin{align}
\mathcal{E}^{(3)}=\int _{\tau}^{\epsilon +\tau}\mathcal{U}\sqrt{h}d^{3}x= \frac{7\kappa ^{2}}{16} \int _{\tau}^{\epsilon +\tau}\frac{1}{\left[\frac{3}{2}(R-\tau)\right]^{10/3}}\left(R-\tau \right)dR \sim \frac{1}{\epsilon ^{4/3}}
\end{align}
which still diverges in the $\epsilon \rightarrow 0$ limit but only as $\epsilon ^{-4/3}$. Thus, in the Schwarzschild spacetime region (outside the dust sphere), close to the  singularity, the energy density due to scalar field tends to arbitrarily high value. The volume factor does help in $(1+3)$ but not completely. 

Thus, on the whole, the results suggest that the back-reaction is important \emph{both} in the outside Schwarzschild regime as well as inside the dust sphere. This has the potential of changing the geometrical structure near the singularity both inside and outside the matter, due to the backreaction. 
 
We will end this section with a few comments on how our results compare with those obtained in some previous attempts \cite{Howard1984, Jensen1989, Anderson1993, Anderson1995}. In most of these studies, approximate expressions for $\langle T_{a b}\rangle$ have been obtained using a fourth order WKB expansion for the field modes to get unrenormalized $\langle T_{a b}\rangle$ and then eliminating DeWitt-Schwinger counterterms \cite{Christensen1976} to get the renormalized value. All these approximate results (which includes both massless and massive fields) for vacuum expectation value of the stress tensor has been obtained outside the event horizon. In \cite{Hiscock1997} the approximate renormalized stress tensor in four-dimensional spacetime was obtained in the interior of the event horizon. Our results follow from the s-wave approximation \cite{Fabbri2005} in which the dominant contribution to the Hawking effect comes from the monopole term ($\ell=0$) in the multipole expansion. This is also well justified 
since we are using the radial observers to foliate our spacetime. 

As we are interested in the region near the singularity, we will consider dominant terms in the observables when the limit $r\rightarrow 0$ is taken. If we calculate the energy density, i.e., $\langle T_{ab}\rangle u^{a}u^{b}$ for radially in-falling observer using the approximate stress energy tensor given in the ref. \cite{Hiscock1997}, it diverges as the singularity is approached (which has been pointed out in ref. \cite{Hiscock1997} itself), but more importantly as $\sim 1/r^{5}$. This is exactly the divergence we have obtained through our analysis as well. Hence, the energy density expressions obtained under s-wave approximation and energy density calculated using approximate renormalized stress energy tensor given in \cite{Hiscock1997} have similar divergent behavior near the singularity. Moreover, the renormalized energy momentum tensor obtained in \cite{Hiscock1997} includes curvature coupled scalar field as well. Hence the divergent nature of the energy density is a generic feature independent of 
coupling with curvature.

Using perturbations around the Schwarzschild solution and treating the renormalized energy momentum tensor as a source for this perturbation, it is seen in ref. \cite{Hiscock1997} that Kretschmann scalar diverges more rapidly for the perturbation compared to the classical Kretschmann scalar. This key result also follows and gets verified in our analysis. Along with these, a related fact that curvature for the perturbation grows more rapidly than the background geometry itself is also consistent with our results. 

\section{Energy density and flux on events along specific null rays}
\label{Paper3:SecNULLALL}

Combining all the results for the static and the radially in-falling observers 
we can describe the variation of the energy density and flux on events along the null rays introduced previously. 
For the events along the null rays inside the horizon the best probes are 
the radially in-falling observers, as they pierce through every spacetime event in the 
inside region i.e. both in the regions A and B [see \fig{Penrose}]. Thus each 
radially in-falling observer will intersect the null rays at one unique spacetime event. 
As the initial radius of the radially in-falling observer is varied, the trajectory will intersect the null 
ray at a different but still unique spacetime point. All these null rays in the outgoing mode 
have constant $V^{-}$ and hence the only parameter that varies along the null rays is $V^{+}$. For the events along the null rays outside the horizon  we can determine the energy density and flux by 
using either the freely falling observers or the static observers. \\
\\
We summarize below all the results obtained for these three rays from our analysis. 

\begin{figure*}
\begin{center}

\includegraphics[height=2in, width=3in]{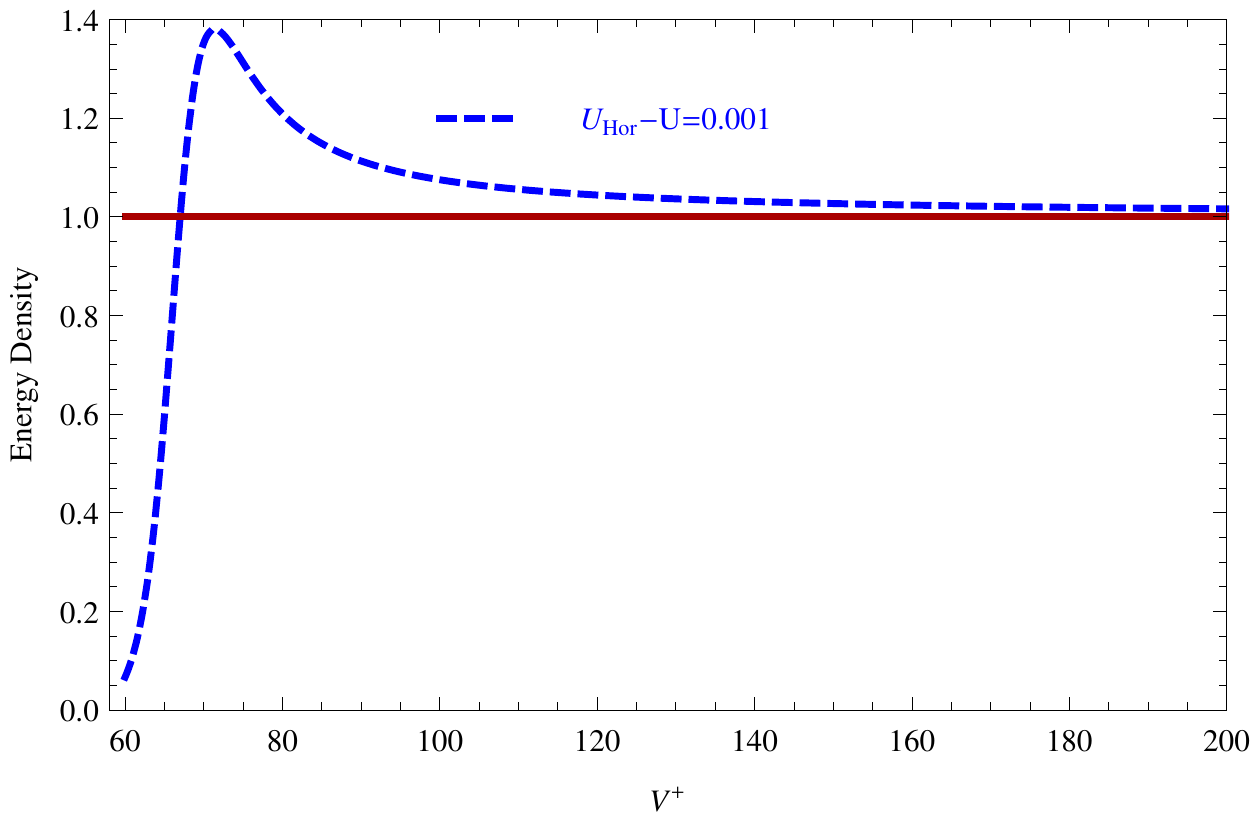}~~~
\includegraphics[height=2in, width=3in]{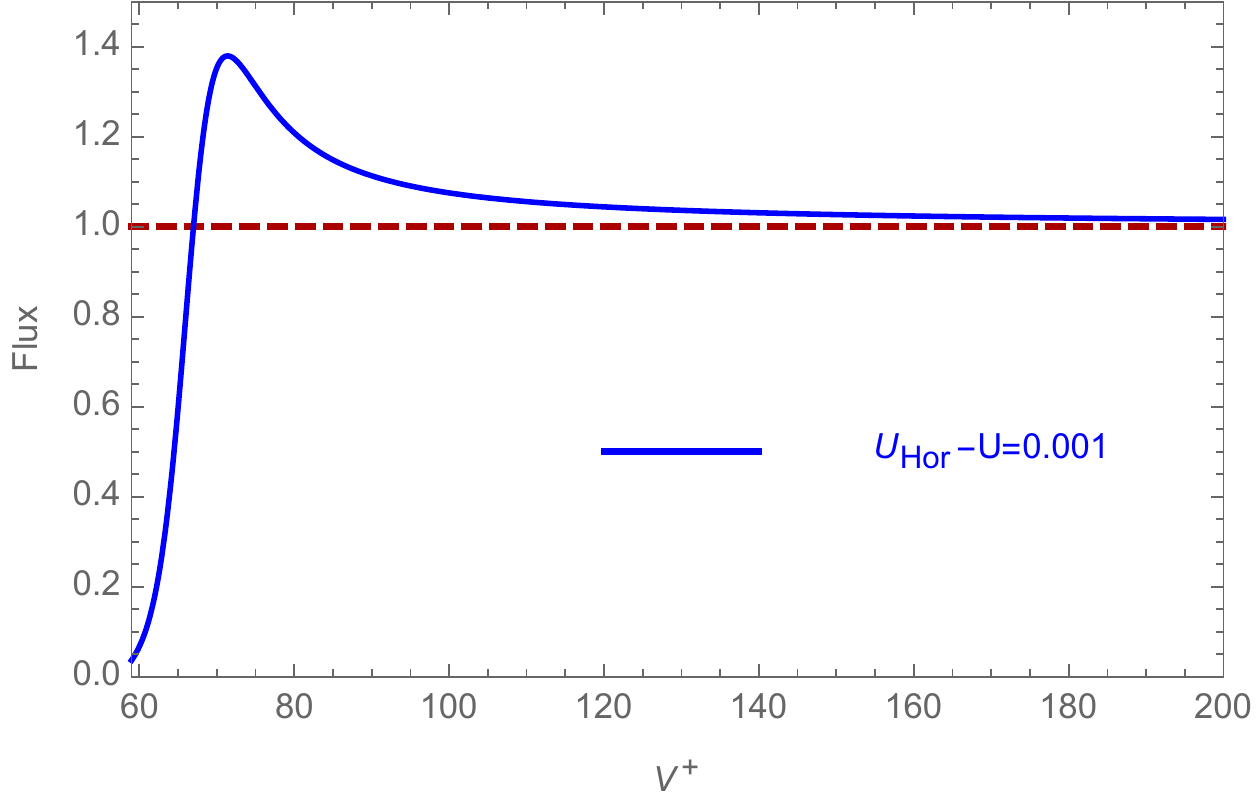}

\caption{ The variation of the energy density and flux on the events along a null ray as a function of
time have been shown. Both the curves saturate at late 
time to the Hawking radiation values implying that all the
events on $\mathcal{J}^{+}$ do receive the standard Hawking flux. The rise of both the flux and energy density at lower values of $V^{+}$ 
corresponds to the near horizon behaviour of these invariant observables. See text for more discussion.} 
\label{Paper3:FigNull01}

\end{center}
\end{figure*}

\paragraph*{Events along the null ray just outside the horizon:} Let us start the discussion by considering the behaviour 
of the energy density and flux on the events along the ray which straddles the horizon just outside and  
ultimately escapes to future null infinity $\mathcal{J}^{+}$. (This is the ray shown in top-right
figure of Fig. \ref{Penrose}.) As we have discussed earlier, there is nothing peculiar happening in regions C and D, and hence we will not bother to discuss those regions. After its reflection at $r=0$ 
the energy density rises, reaches a maxima and then drops back to the Hawking value as the 
asymptotic infinity is approached [see the left plot in \fig{Paper3:FigNull01}]. 
Similar behaviour is exhibited by the flux, 
which also reaches a maxima and then decreases ultimately reaching the Hawking value as the asymptotic limit is approached. 

Hence we conclude that, on the events along the null ray just outside the horizon, nothing strange happens and the results reproduce the standard black hole radiation, known in the literature.
\paragraph*{Events along the null ray just inside the horizon:} The second null ray passes through the events which are inside the event horizon (See the bottom-left figure in \fig{Penrose}). Initially this ray shares the same geometry as the previous one and hence has almost the same  energy density and flux. It also straddles the horizon but--- being inside the horizon--- its ultimate fate is sealed; it has to hit  the singularity in finite time [see \fig{Penrose}]. Thus even though the initial events along these two rays share similar physical conditions, the final regions of spacetime encountered by these two rays are very different. One ends at the future null infinity as described earlier, while the other one ends at the singularity. 

\begin{figure*}
\begin{center}

\includegraphics[height=2in, width=3in]{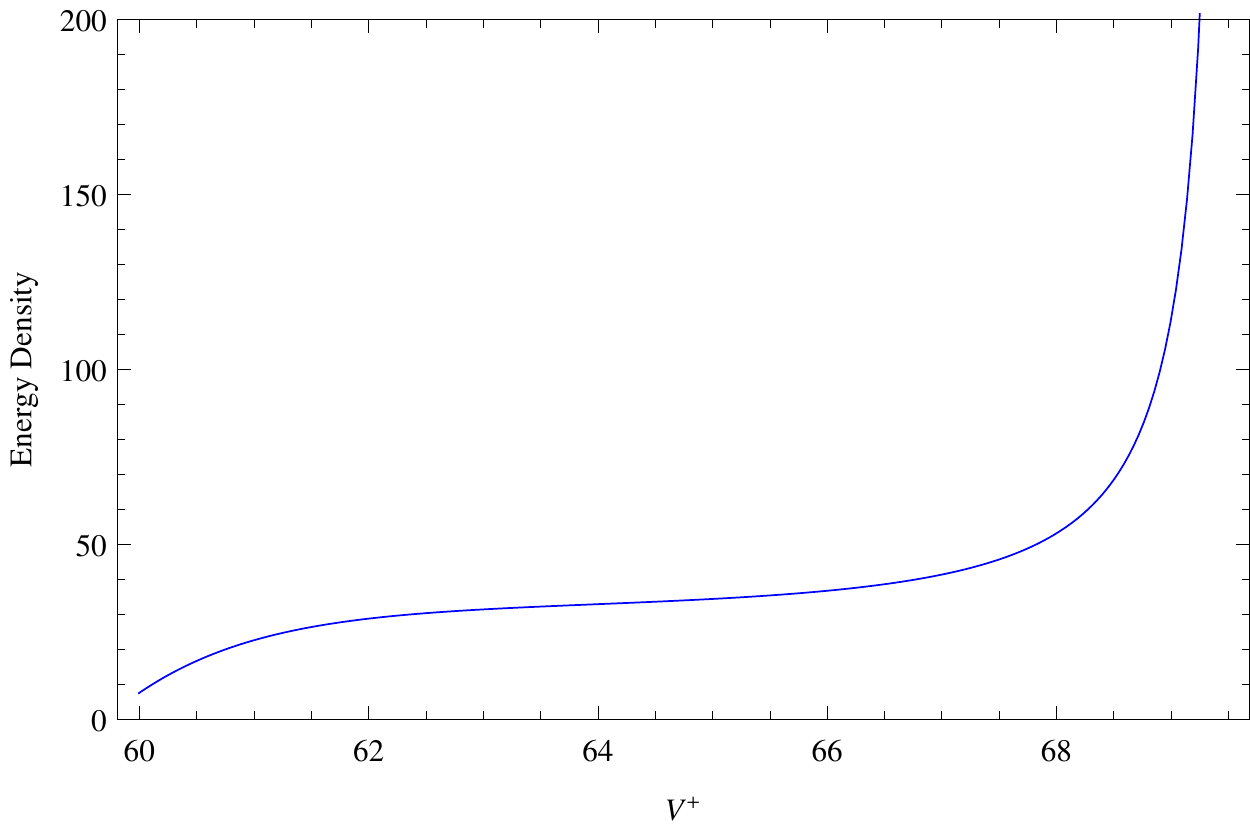}~~~
\includegraphics[height=2in, width=3in]{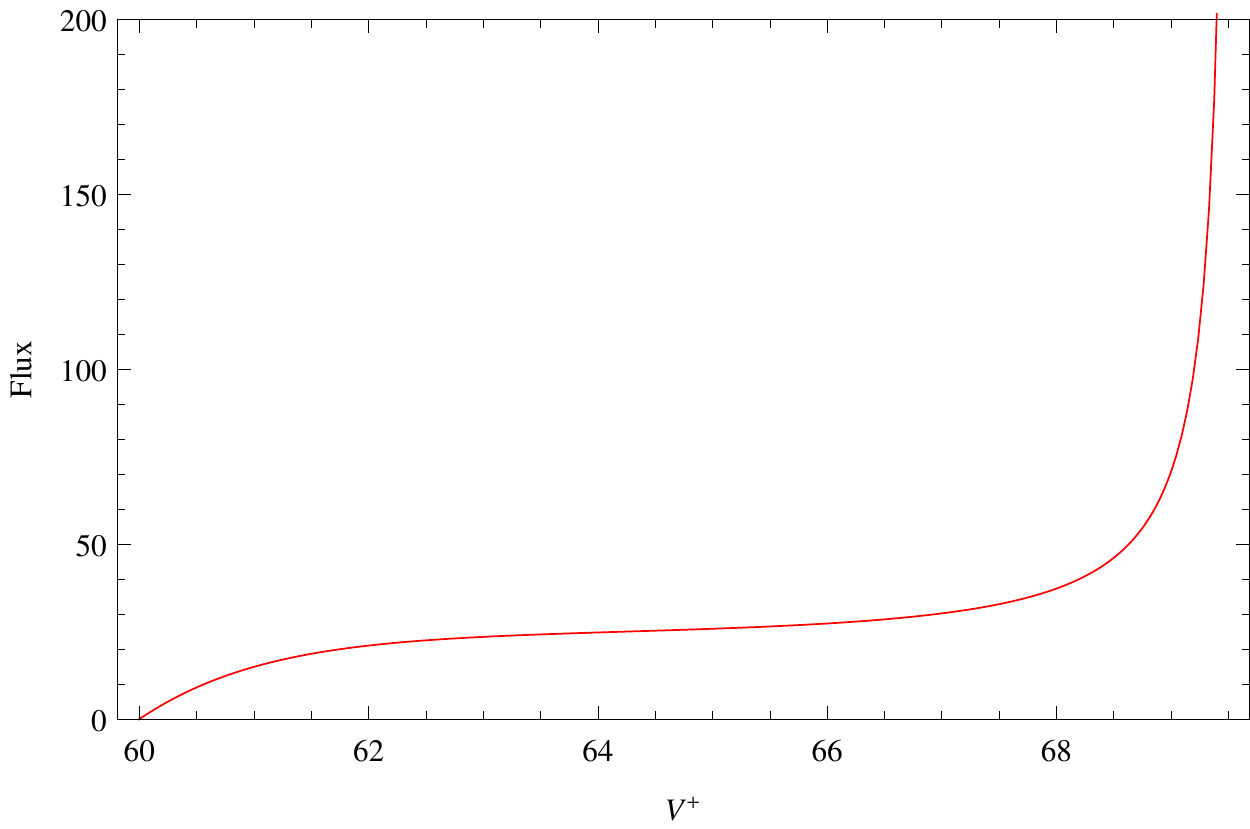}\\

\caption{The energy density and flux along null rays very near --- but inside --- the horizon. These rays come out of the dust sphere after the surface of the dust sphere has crossed the event horizon. Since all these null rays hit the singularity in a finite proper time the energy density and flux diverge. See text for detailed discussion.}
\label{Paper3:FigNull02}

\end{center}
\end{figure*}

The energy density and flux in the present case can be calculated using the radially in-falling observers. For different radially in-falling observers (parametrized by the radius from which the free fall starts) the null ray cuts these observers at unique and distinct points. Thus calculating the value of the invariant observables at every point enables us to obtain their variation on the events along the null ray. The energy density along this null ray starts to rise and ultimately diverges as the singularity is approached [see the first figure of \fig{Paper3:FigNull02}]. The same feature is also seen in the flux as measured along this null ray. 

\begin{figure*}
\begin{center}

\includegraphics[height=3.5in, width=5in]{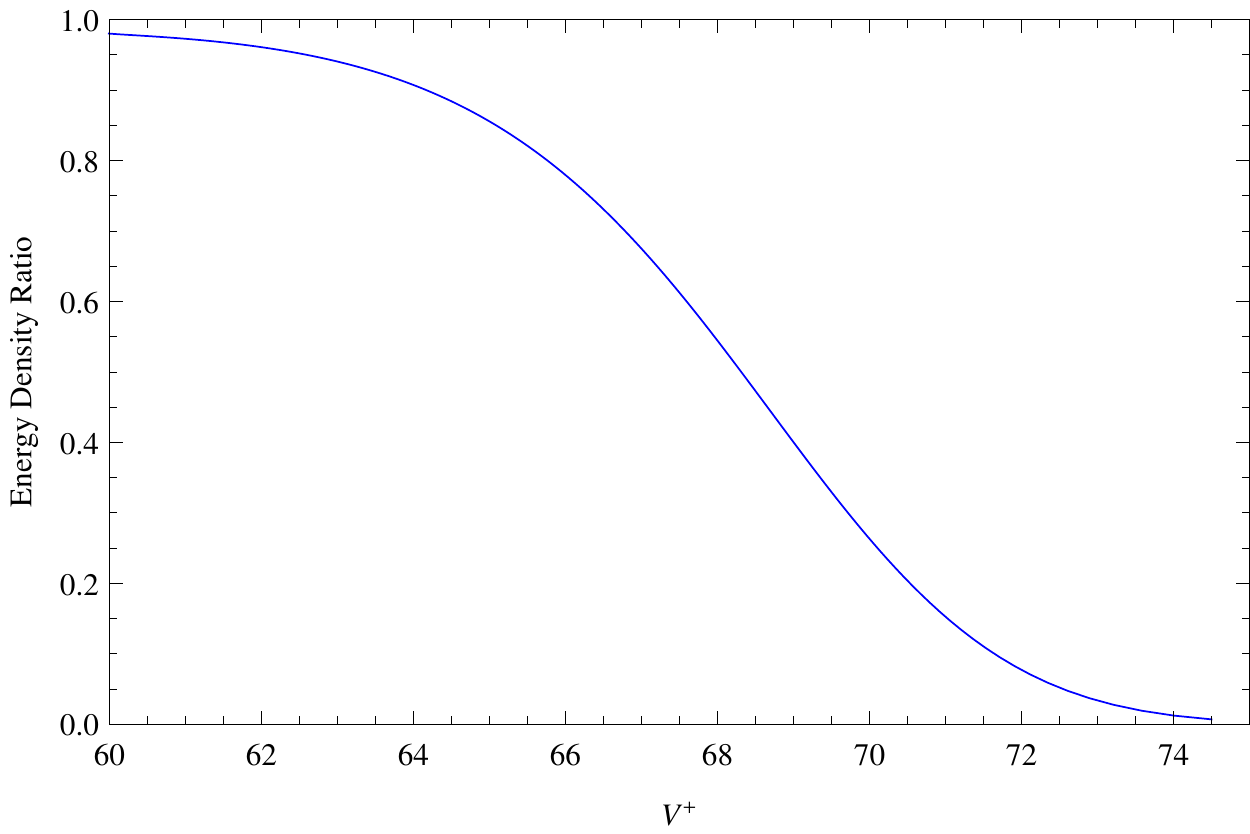}

\caption{The ratio of the two energy densities along the two 
null rays straddling the event horizon --- one just one outside, while another just inside. 
As the inside ray approaches the singularity 
this ratio reduces to zero since energy density along the inside ray diverges. 
However for the rays near the event horizon --- but outside --- the energy density is finite, 
making the ratio  vanish at late times. 
For two null rays with almost identical $V^{+}$ at the time of reflection at $r=0$  
the energy density turns out to have similar values; it can be seen  from the figure  that the ratio
at the beginning is almost equal to unity.}
\label{Paper3:FigNull03}

\end{center}
\end{figure*}

Thus we can conclude that the final phase of the journey for this null ray shows significant differences compared to the one discussed earlier. For events along this ray close to the singularity, both the invariant observables diverge, in striking contrast to the events along the previous null ray, which ultimately leads to  the Hawking flux [see \fig{Paper3:FigNull03}]. Naively speaking this makes the back reaction effects quite significant close to the singularity. However, as we said before, such a divergence in the energy \textit{density} can be compensated by the shrinking 3-volume in a region close to the singularity. So we needed to compute the total energy inside a small volume before we can conclude about the divergence. This question has already been addressed in \sect{RadFallOut}. The final conclusion to be drawn is that these quantities do exhibit a divergence even in (1+3) spacetime dimensions.

\paragraph*{Events along the null ray inside the dust sphere:} This is the last situation we need to consider. This null ray stays mostly inside the dust sphere and also hits the singularity while it is still inside. On the events along this null ray we can make a direct comparison of the energy density of the quantum field with that of the dust sphere itself. As it moves towards the event horizon the energy density again rises, reaches a maxima and then diverges as the ray hits the singularity [see \fig{Penrose}]. The same behaviour was also noted earlier for radially in-falling observers inside the dust sphere. As these observers are used to define the invariant observables along the null rays they exhibit similar behaviour. 
\begin{figure*}
\begin{center}

\includegraphics[height=2in, width=3in]{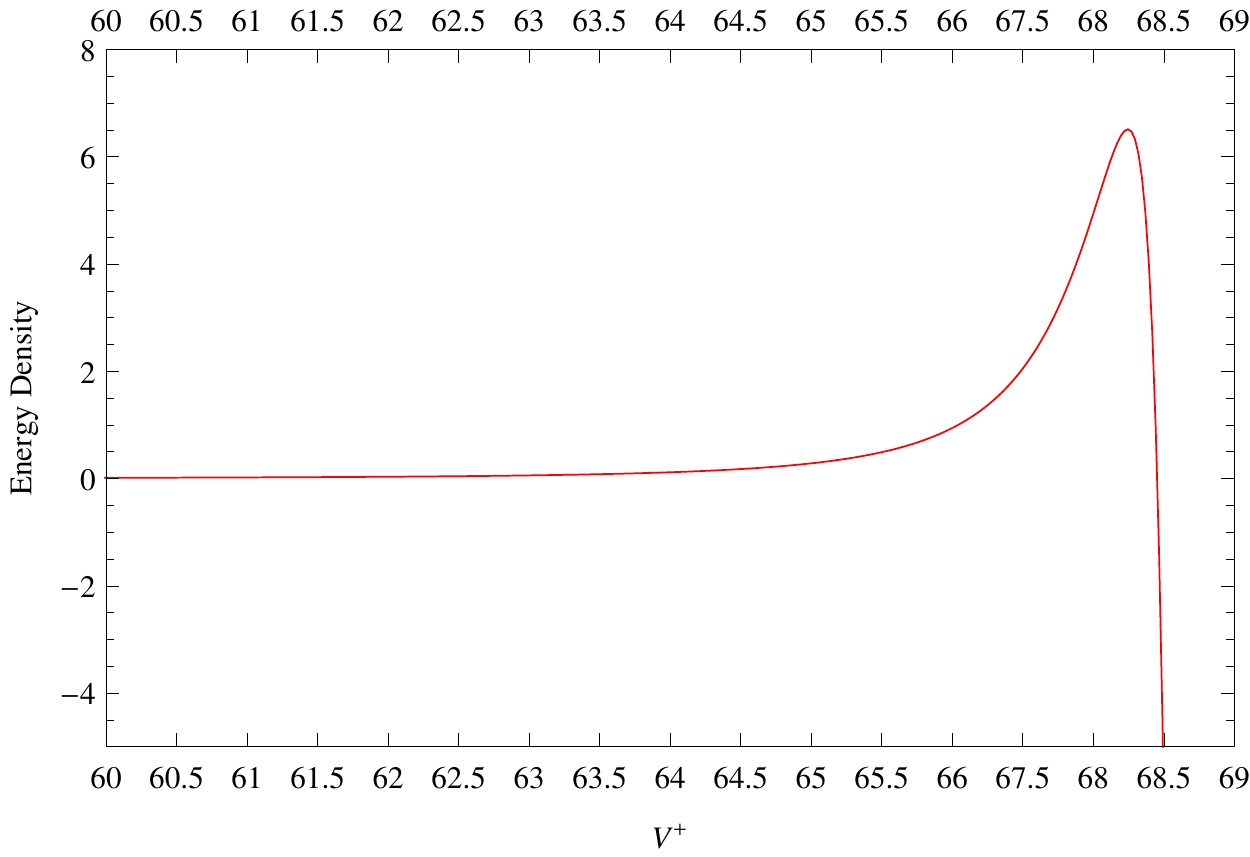}~~~
\includegraphics[height=2in, width=3in]{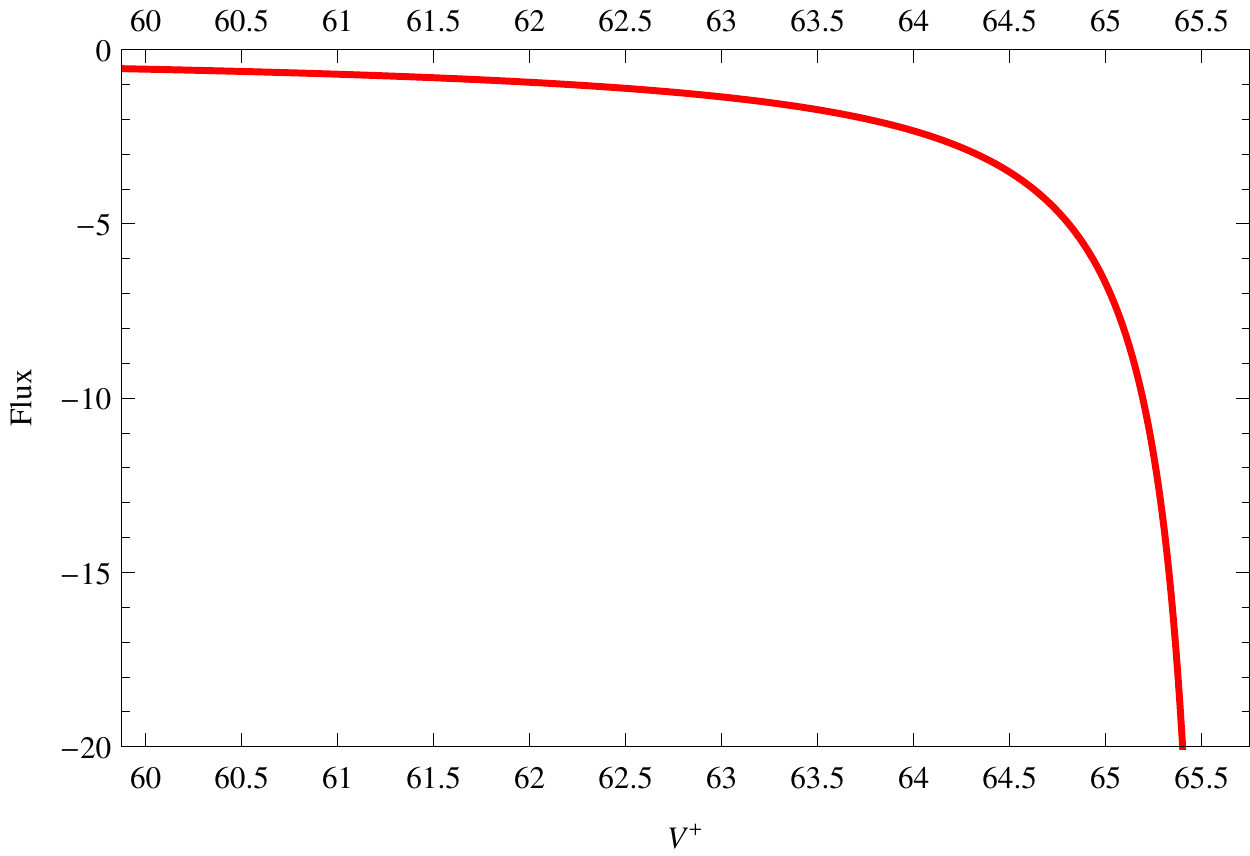}

\caption{ Behaviour of $\mathcal{U}$ and $\mathcal{F}$ at events along the null rays hitting the singularity while they are within the dust sphere. For these rays the value of the $U$ coordinate is greater than the value of the $U$ coordinate at which the dust ball hits the singularity. The null ray initially shows a growth  as  seen for the radially in-falling observers inside the dust sphere. After reaching a maximum the energy density ultimately decreases, finally it diverges as it hits the singularity. Identical divergence can also be seen for the flux. See text for detailed discussion.}
\label{Paper3:FigNull04}

\end{center}
\end{figure*}

In this case as well the ratio $\mathcal{U}/\rho$ diverges as the singularity is approached, i.e. the ratio tends to arbitrarily high value as the null ray reaches arbitrarily close to the singularity. The same result was also obtained earlier in the case of radially in-falling observers inside the dust sphere. Even though the above conclusions were drawn from (1+1) spacetime dimensions they carry forward to (1+3) spacetime dimensions as well, as shown in \sect{RadFallIn}. Thus we conclude that the respective energy density for the scalar field and its ratio with dust energy density diverges in (1+1) as the singularity is approached. In (1+3), even though the energy for the scalar field does not diverge due to the volume factor in (1+3) spacetime dimensions, the ratio $\mathcal{E}^{(3)}/E^{(3)}$ diverges. Thus the backreaction \emph{is} important within the dust sphere as well. As we saw earlier, this is the case at events in region B close to the singularity, where the back reaction effects are significant.

\section{Effective Temperature measured by Detectors }
\label{effectemp}

\subsection{Effective Temperature as a measure of detector response}\label{Paper3:SecTemp}

We believe, for reasons described in \sect{Paper3:SecSET}, that the regularised stress-tensor expectation value is the appropriate quantity to study in our case. Neverthless, for the sake of completeness and to compare our results with those obtained earlier in the literature, we will also consider the detector response in this Section. 

Let $\gamma (\tau)$ be the trajectory of an asymptotic stationary detector expressed in terms of its proper time $\tau$.  Then, under certain conditions, one can associate \cite{Smerlak2013,Barbado2011,Barcelo2011} a temperature $T_{-}$  with this detector by:
\begin{equation}\label{Paper3:Eq24}
T_{-}=\frac{1}{2\pi}\left \vert \frac{\ddot{V}^{-}}{\dot{V}^{-}}\right \vert
\end{equation}
(One can also associate a second  temperature $T_{+}$ in an analogous fashion; but we will not need it for our discussion.) 
The usefulness of the above definition arises from the fact that it can be applied to non-stationary, non-asymptotic observers as well. It can be easily verified \cite{Smerlak2013} that $T_{-}(\tau)$ indeed leads to the Tolman redshifted temperature \cite{Tolman1930} as measured by static Unruh-DeWitt detectors. The approximate constancy of $T_{-}$ can be expressed through the \emph{adiabaticity condition} \cite{Barbado2011,Barcelo2011} which can be expressed as:
\begin{equation}\label{Paper3:Eq25}
\eta _{-}\equiv \left \vert \frac{\dot{T}_{-}}{T_{-}} \right \vert \ll 1
\end{equation}
This condition allows us to consider the response of the Unruh-DeWitt detectors in a straightforward way, by evaluating $T_{-}(\tau)$. Hence with every trajectory we can introduce a temperature $T_{-}(\tau)$ along with its adiabaticity parameter $\eta _{-}$. This setup will provide another handle on the quantum field theory of the scalar field in the collapsing background geometry.

\subsection{Static Detectors in region C}

\begin{figure*}
\begin{center}

\includegraphics[height=2in, width=3in]{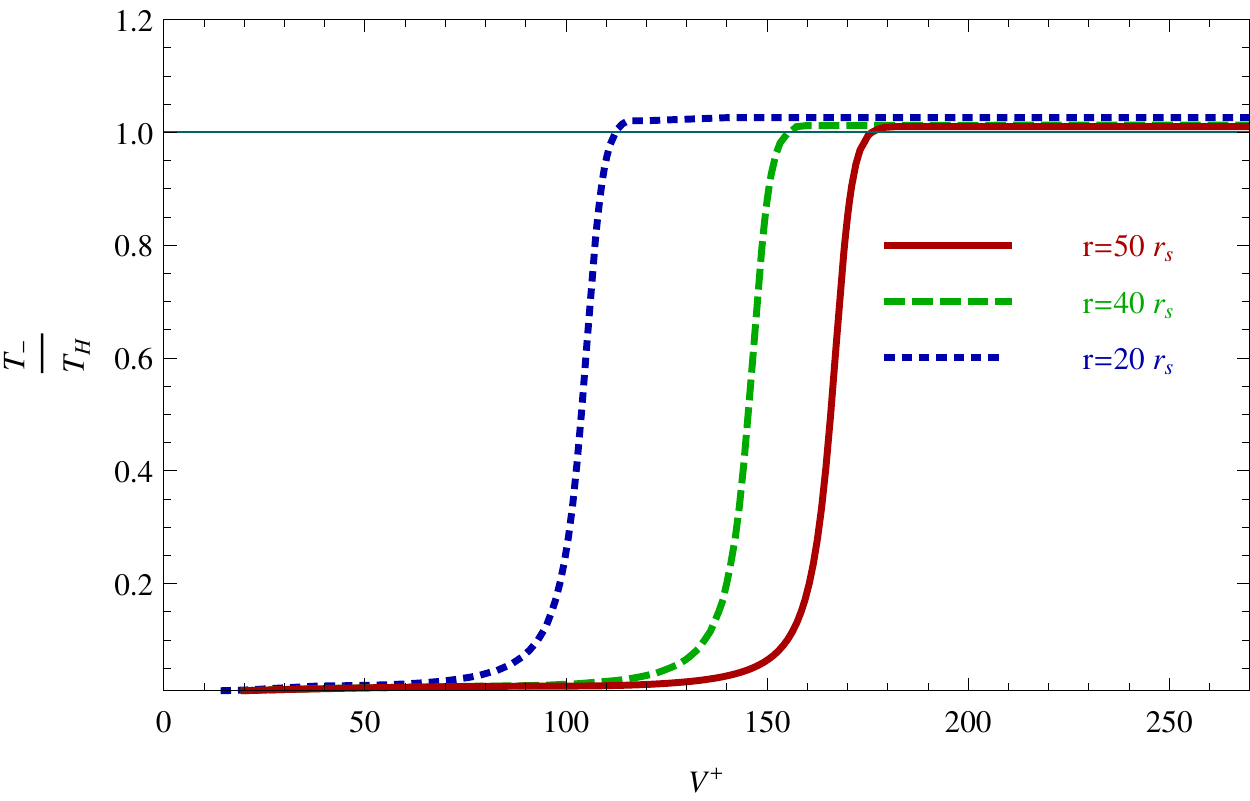}~~
\includegraphics[height=2in, width=3in]{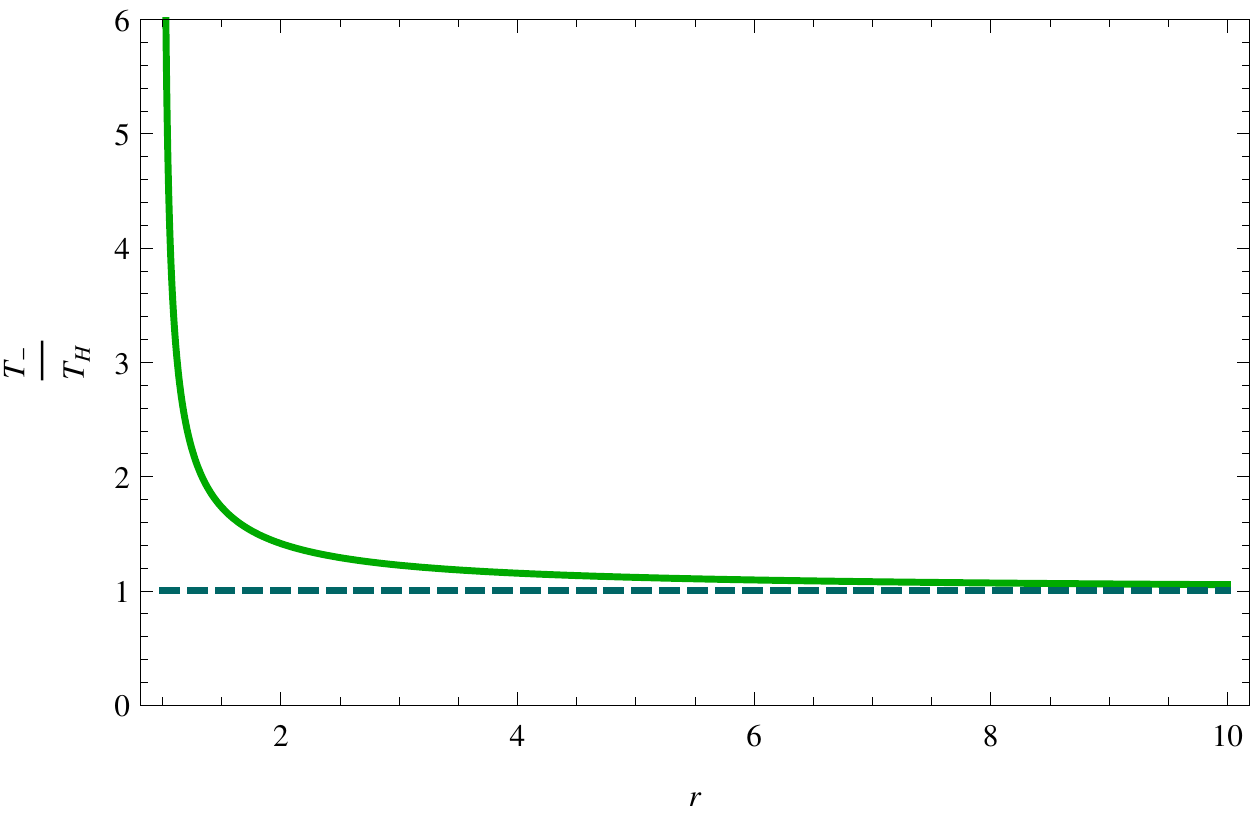}\\

\caption{The effective temperature 
  $T_{-}$ reaches the Tolman redshifted 
Hawking temperature  $T_{H}$ at late $V^+$ (left figure). In the near horizon regime the 
effective temperature $T_{-}$ is positive and diverges as the horizon is approached (right figure).}
\label{Paper3:FigSTATTemp01}
\end{center}
\end{figure*}

For a static detector the effective temperature corresponding to $V^{-}$ exists which, in the late time limit, ($U\rightarrow \pi -3\chi _{0}$) reduces to: 
\begin{equation}\label{Paper3:Eq30}
T_{-}^{\textrm{late}}=\frac{1}{4\pi}\sqrt{\frac{r}{r-1}}=T_{H}\sqrt{\frac{r}{r-1}}
\end{equation}
(the general expression is given in \eq{Paper3:Eq29} in the appendix.) This is precisely the Tolman redshifted temperature, showing the validity and use of this effective temperature formalism. The variation of the effective temperature with $V^{+}$, proportional to the proper time of the static detector has been shown in \fig{Paper3:FigSTATTemp01}. We have also illustrated the variation of the effective temperature at late time with radii, showing the divergence as the horizon is approached. 
Incidentally, the detector temperature allows us to study another question which is of interest by itself. Consider a collapsing structure which --- unlike the pressure-free dust studied so far --- exhibits the following behaviour. The dust sphere starts to collapse from certain radius, continues to collapse untill it reaches $r=2M+\epsilon$ (with $\epsilon\ll 2M$) and then stops collapsing due to, say, internal dynamics and becomes static. What kind of radiation will be detected by a static detector at large distances?  Previous studies, based on quantum field theory\cite{Paranjape2009}, has shown that the collapsing body (i) will emit radiation closely resembling the Hawking effect during the collapsing phase and (ii) the effective temperature will drop down to zero at late times. It worthwhile to study this situation using the detector response.  

\begin{figure*}
\begin{center}

\includegraphics[height=2in, width=3in]{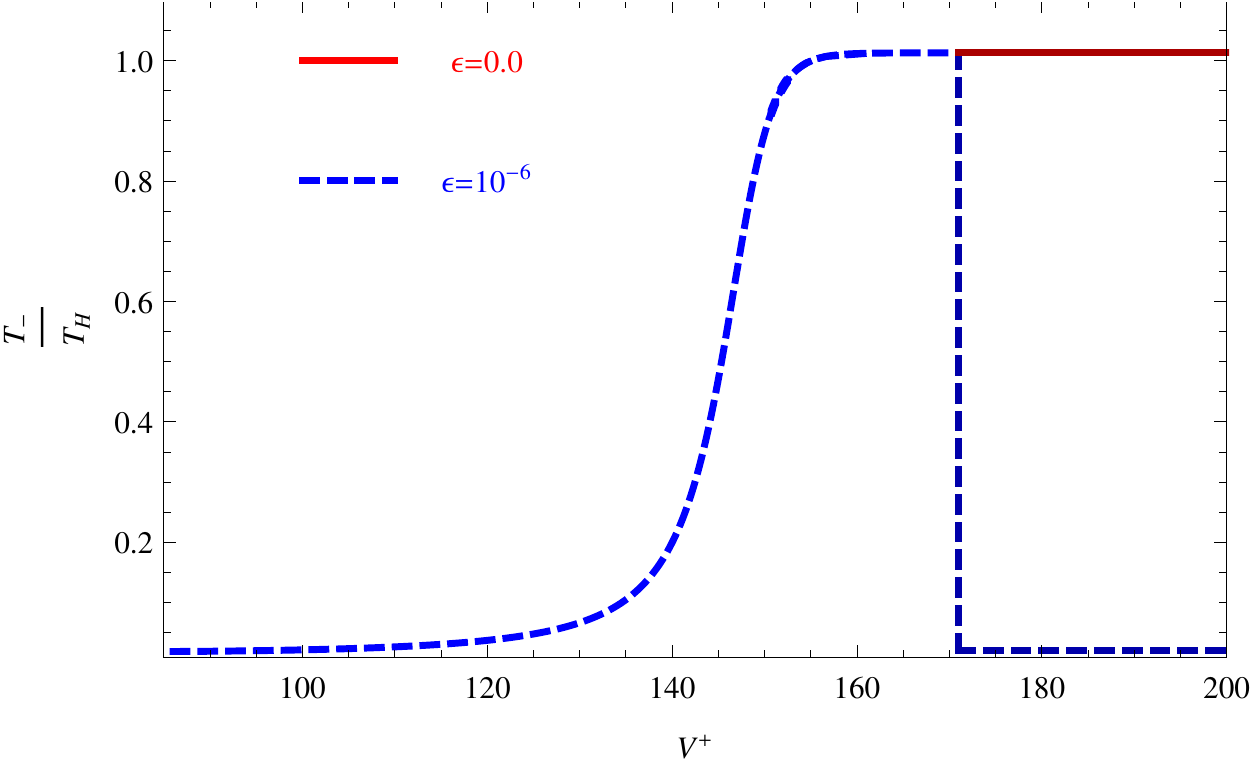}~~
\includegraphics[height=2in, width=3in]{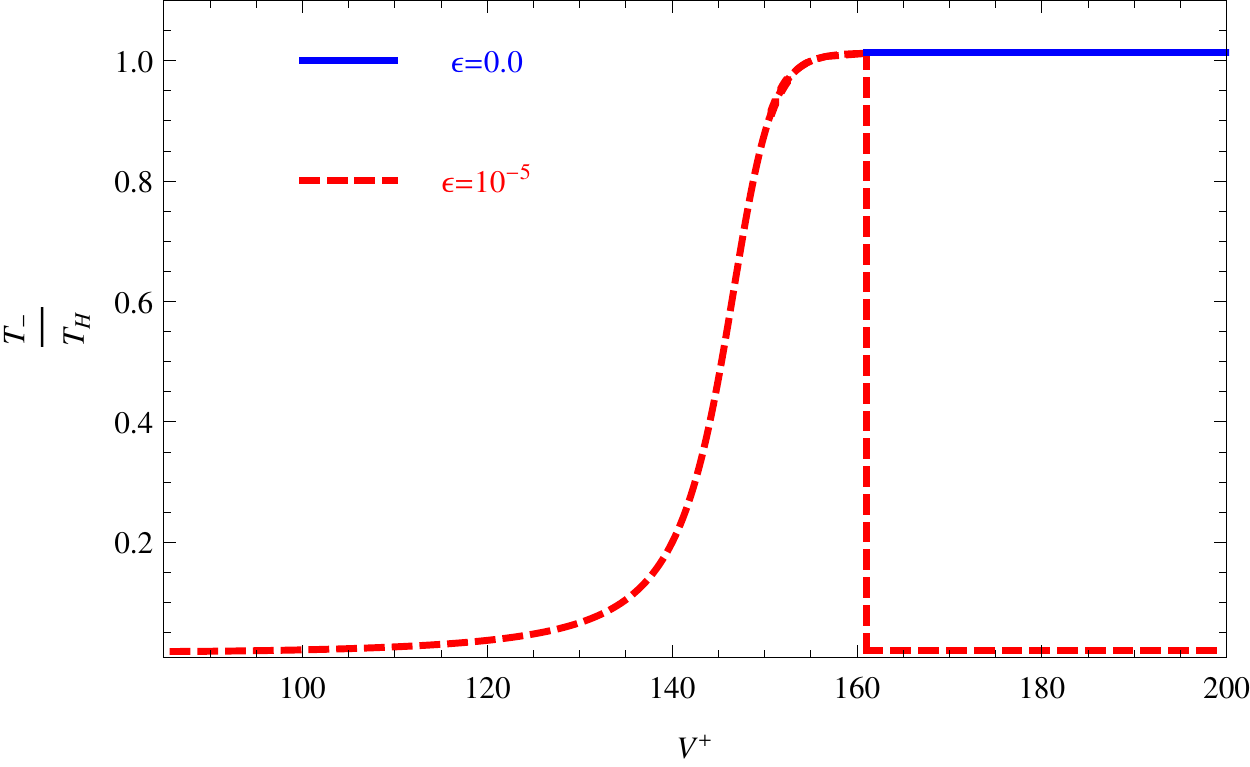}

\caption{The variation of the effective temperature with $V^{+}$ as the surface of the dust sphere approaches the horizon. 
If, for some reason, the collapse is halted at $r=2M+\epsilon$ (before forming the horizon)the effective temperature --- which initially raises towards the standard Hawking value --- drops to zero when the dust sphere becomes static. 
This situation is shown for two values of the  final radii in the two 
plots.}
\label{Paper3:FigSTATTemp02}

\end{center}
\end{figure*}

We will consider the detector response i.e. effective temperature measured by static detectors in this collapse scenario. The relevant collapsing geometry can be easily achieved by applying the same equations as given by \eq{Paper3:Eq9} with $\eta$ being restricted to the range: $0\leq \eta \leq \pi -2\chi _{0}-2\epsilon$. Here $\epsilon$ is connected to the quantity $r_{stop}-r_{s}$ i.e. the radial distance from the horizon at which the collapse stops. This relation can be easily obtained as: $r_{stop}-r_{s}=2\left(\cot \chi _{0}\right)\epsilon$. Hence, given the value of $\epsilon$, we can obtain the effective temperature at all times. We find it to be nonzero and approaching the saturated Tolman redshifted Hawking value during the collapse. While after the collapse stops there is no formation of trapped region and thus the effective temperature drops to zero. This situation is being depicted in \fig{Paper3:FigSTATTemp02} for two different choices of $\epsilon$. It turns out that as $\epsilon$ decreases 
the effective temperature more closely resembles the Hawking temperature. As $\epsilon \rightarrow 0$, we recover the usual Hawking evaporation result. 

Let us now return to our main theme. Having discussed the energy density, flux and effective temperature for 
static observers, we will now proceed to determine the same for radially in-falling 
observers, both inside the dust sphere and outside. 

\subsection{Radially In-falling Detectors: Inside regions D and A}

\begin{figure*}
\begin{center}

\includegraphics[height=2in, width=3in]{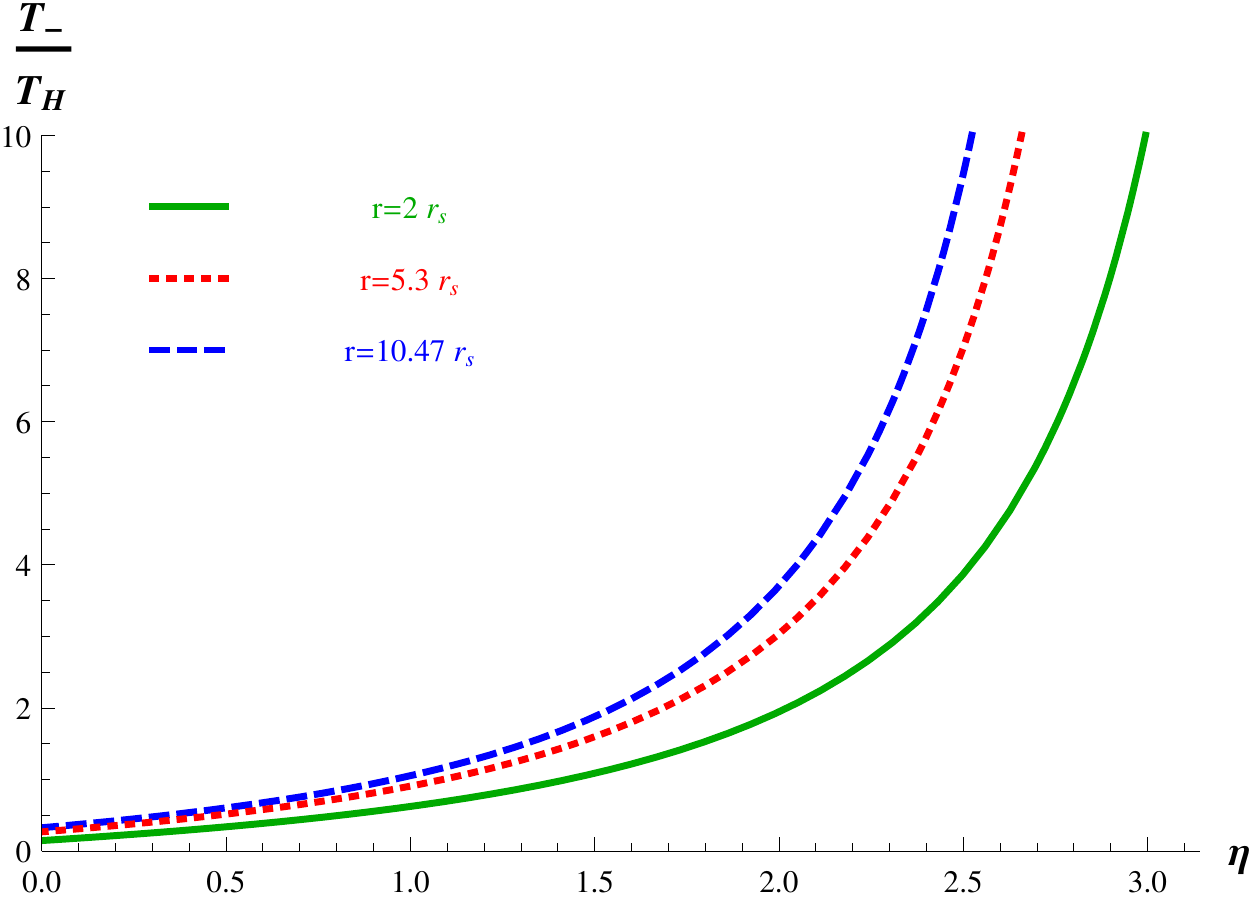}~~

\caption{The effective temperature
$T_{-}$ for radially in-falling detectors comoving with the 
dust sphere and remaining inside it. It is evident that 
the effective temperature diverges as the detector approaches the singularity.}
\label{Paper3:FigInfallTemp}
\end{center}
\end{figure*}

Let us first consider the effective temperature measured by the observers inside the dust sphere. As the spacetime can be globally mapped by the coordinates $(V^{+},V^{-})$, we can follow the same procedure as adopted before to calculate the effective temperature $T_{-}$. For completeness we provide the expression for the effective temperature $T_{-}$ along the trajectory of the radially in-falling detector in \eq{Paper3:Eq32} in the Appendix. The effective temperature depends on the proper time $\eta$ along the trajectory of the radially in-falling observer and is plotted in \fig{Paper3:FigInfallTemp}. It becomes arbitrarily large as $\eta \rightarrow \pi$ i.e. as the radially in-falling detector approaches the singularity arbitrarily close. 

\subsection{Radially In-falling Detectors: Outside regions C and B}

\begin{figure*}
\begin{center}
\includegraphics[height=2in, width=3in]{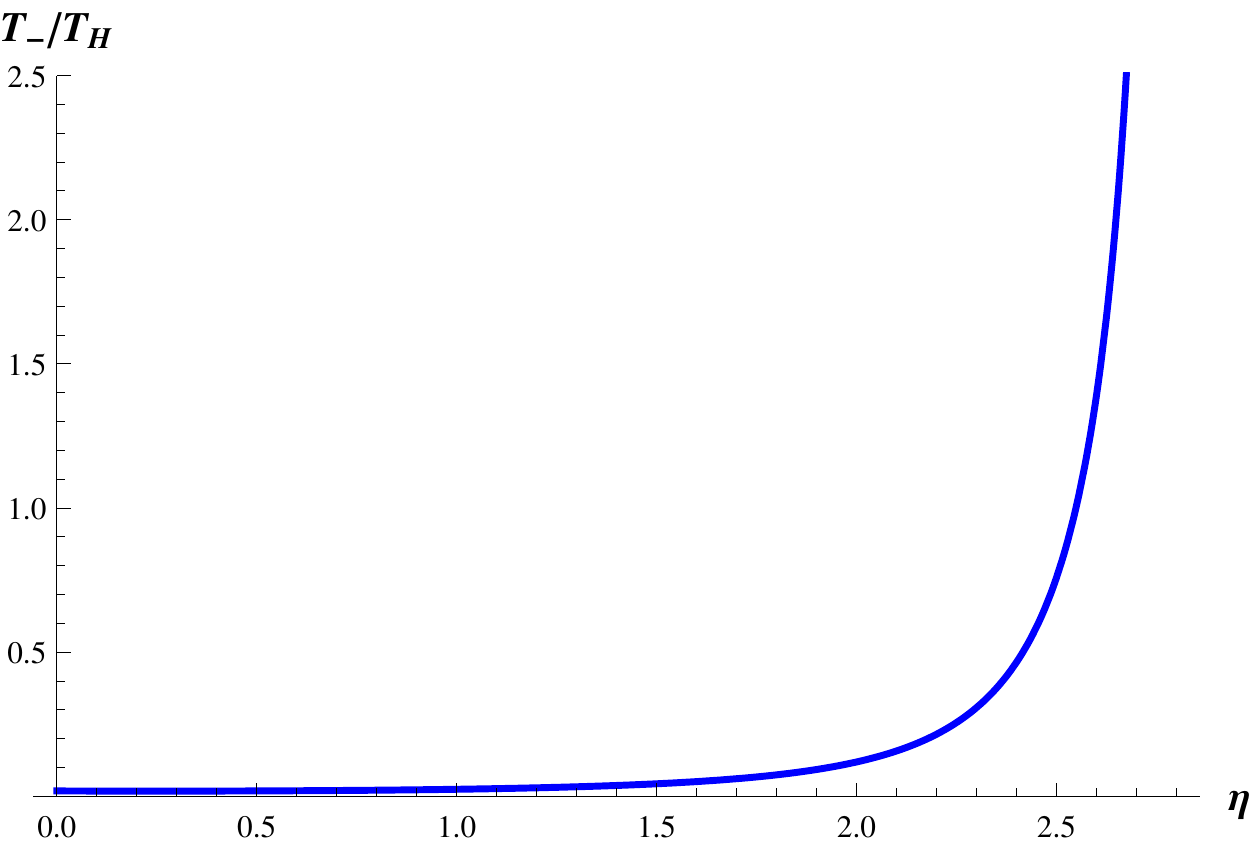}~~~
\includegraphics[height=2in, width=3in]{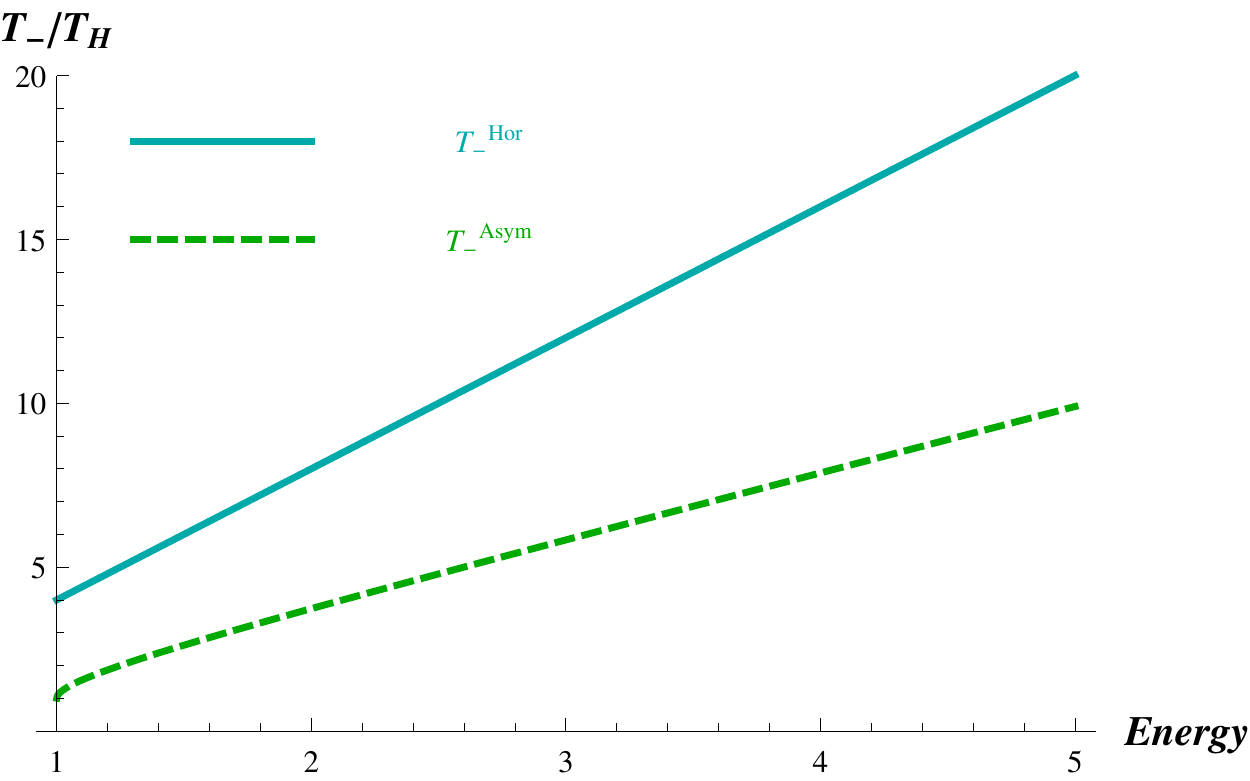}
\caption{The first figure shows  the behaviour 
of $T_{-}$,  normalized to the Hawking value $T_{H}$, 
as the radially in-falling observer approaches the 
singularity. 
The next figure describes the variation of the effective temperature with the energy gap of the 
detector in the asymptotic and near horizon regime. The near horizon value is greater than the 
asymptotic value for unbound trajectories i.e. trajectories with $E \geq 1$.}
\label{Paper3:FigRadInTemp01}
\end{center}
\end{figure*}

Let us now consider the case of Unruh-DeWitt detectors outside the dust sphere moving on a radially in-falling trajectory. The effective temperature they measure along their radially in-falling trajectory is specified by the energy $E$. This is a non-stationary phenomenon, since--- as the detector approaches the singularity the local curvature grows rapidly. Thus this trajectory allows us to probe the time dependent Hawking 
temperature in  detail. The expression is given in \eq{Paper3:Eq41} in the Appendix. 

Here we are interested in two limits: the asymptotic one and the near 
horizon limit. In the asymptotic limit we have $\ddot{u}=0$ and $(r-1)/r=1$, 
such that the effective temperature reduces to:
\begin{equation}\label{Paper3:Eq42}
T_{-}^{\textrm{Asym}}=T_{H}\left(E+\sqrt{E^{2}-1}\right)
\end{equation}
which is consistent with what we expect. For a radially in-falling observer who starts her journey from spatial infinity has energy $E=1$. Then this observer will detect a temperature $T_{-}^{\textrm{Asym}}=T_{H}$. While for observers with $E\ne 1$, the asymptotic temperature would be different from the Hawking value, as obtained earlier in the context of a null collapse in \cite{Smerlak2013}. In this 
regime the adiabatic parameter is negligibly small. 

On the other hand, at horizon crossing, we have the following expressions: 
$\dot{r}=-E, \dot{V}^{+}=(1/2E)$ and $\ddot{r}=-(1/2)$. Then we obtain the following 
expression:
\begin{equation}\label{Paper3:Eq43}
\frac{\ddot{V}^{-}}{\dot{V}^{-}}=\frac{2\ddot{u}-\dot{u}^{2}}{2\dot{u}}
=\lim _{r \rightarrow 1}\frac{4\dot{r}^{2}(r+1)-4r(-\ddot{r}+\dot{r}\dot{v})}{4r\dot{r}}
=-2E
\end{equation}
Thus the near horizon effective temperature is given by :
\begin{equation}\label{Paper3:Eq44}
T_{-}^{\textrm{Hor}}=4E T_{H}
\end{equation}
Thus the effective temperature is not only non zero, but for states
with $E\geq 1$, it exceeds the Hawking value. For an Unruh-DeWitt detector dropped 
from infinity we have $E=1$ and thus the detector will perceive four times 
the Hawking temperature at the horizon crossing. (This result was  obtained 
earlier in Ref. \cite{Smerlak2013,Barbado2011}). However at this stage the adiabaticity 
parameter $\eta _{-}$ has large value, and hence the interpretation of $T_{-}$ as the 
temperature has some ambiguity, and one needs to consider progressively larger frequency 
modes. 

\begin{figure}[h!]
\begin{center}
\includegraphics[scale=0.42]{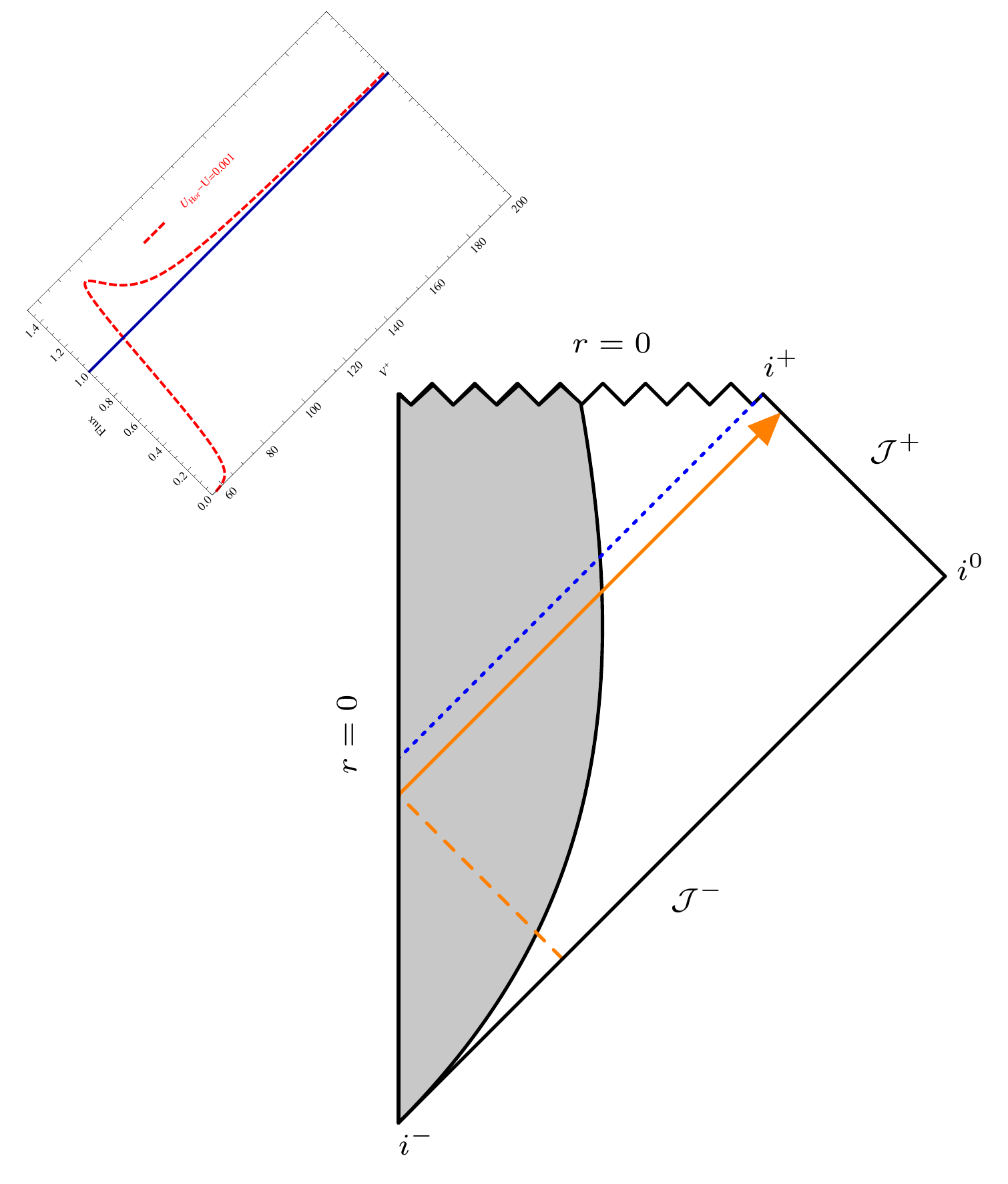}\hspace{15pt}
\includegraphics[scale=0.51]{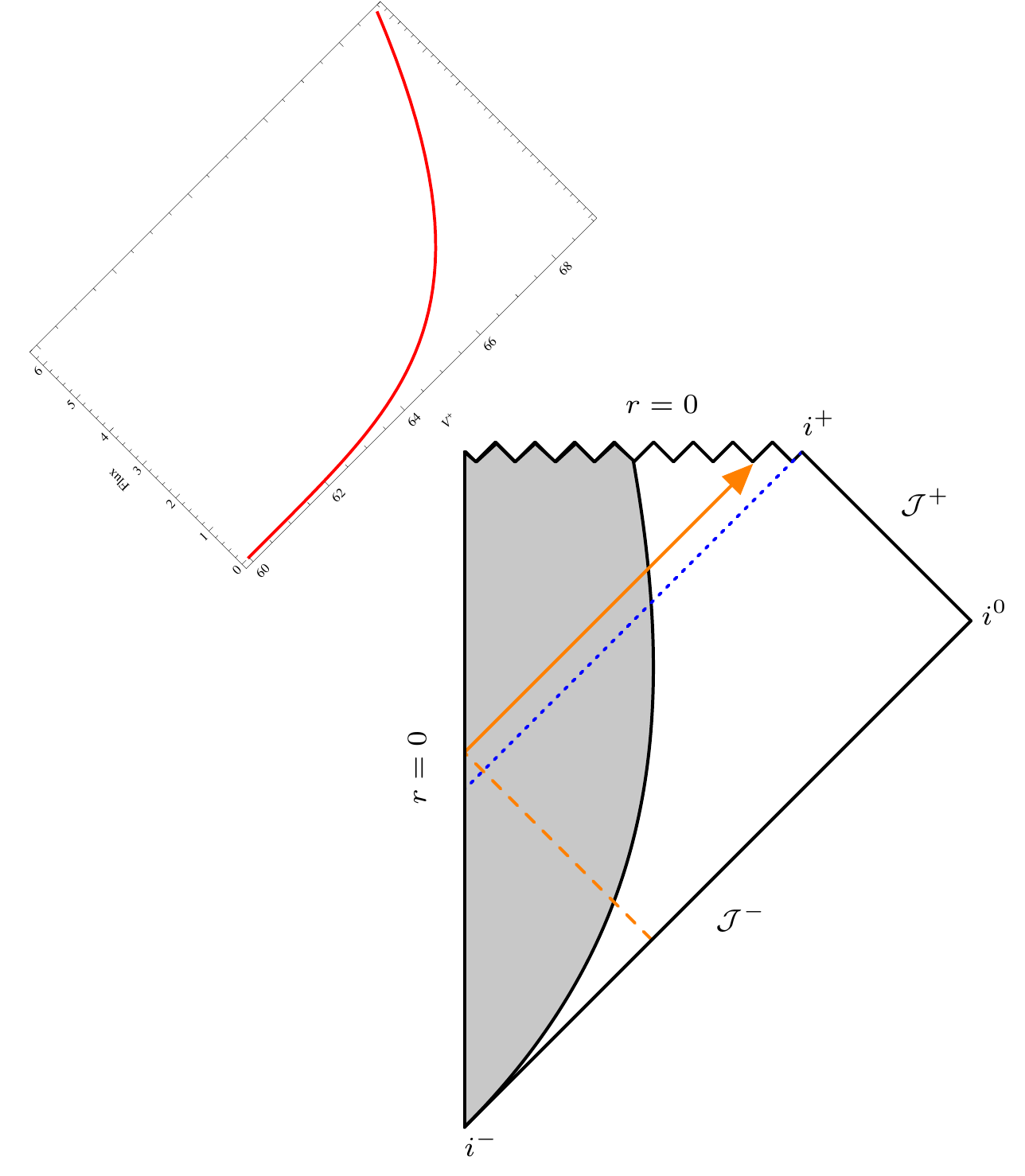}\\
\includegraphics[scale=0.33]{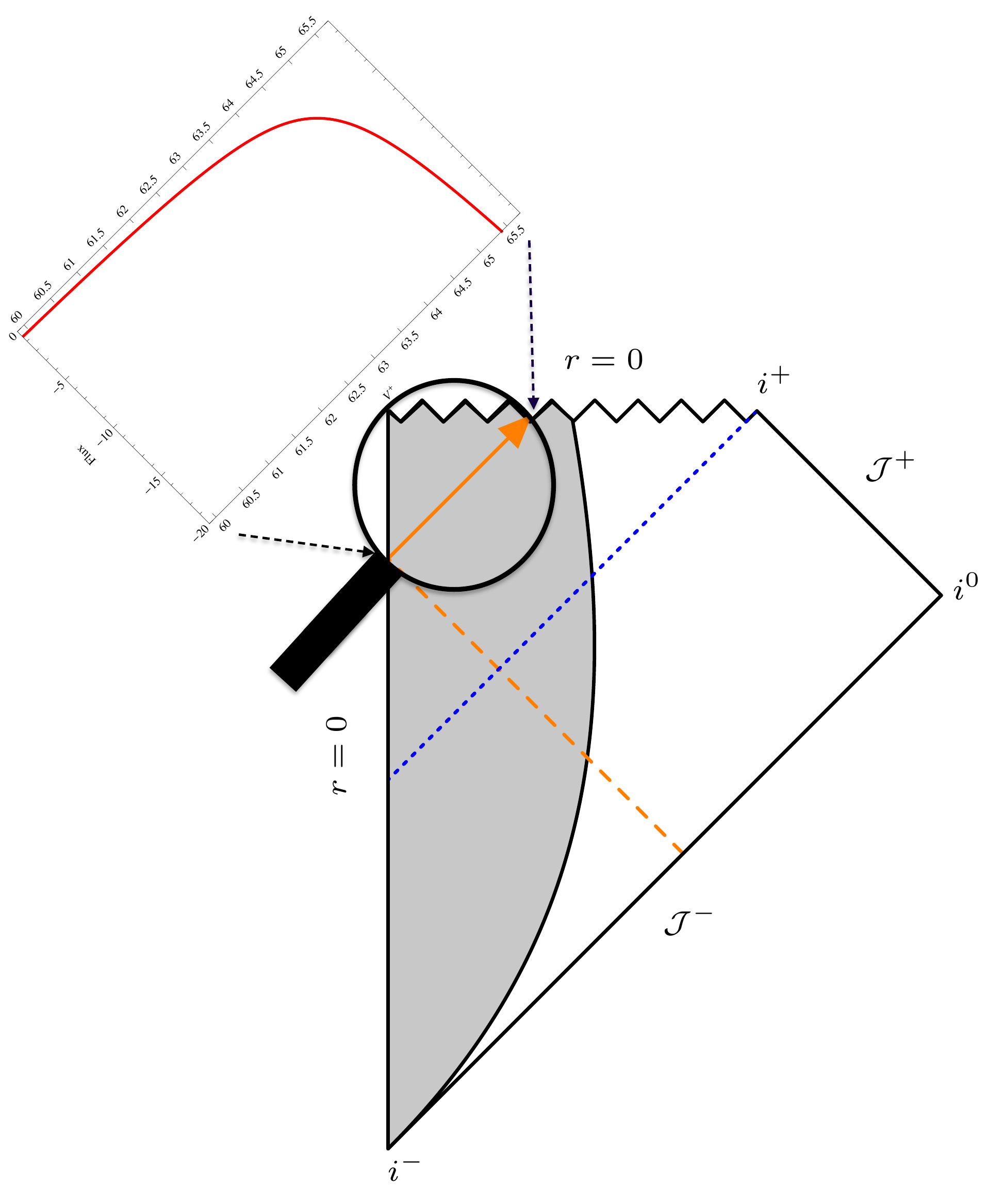}
\end{center}
\caption{ Penrose diagrams (along with the associated graphs) illustrate how the flux along the null rays vary. In the top-left figure we consider a ray straddling the event horizon but remains outside. The flux in the near horizon regime first shows an increment, but soon tends to the Hawking value as the ray approaches the future null infinity. The behaviour of flux along the null ray straddling the event horizon just inside, but remaining \textit{outside} matter at late times is shown in the top-right figure. This null ray hits the singularity in a finite proper time and the flux diverges. The bottom figure shows the divergent nature of the flux for null rays \textit{inside} the dust sphere. The flux diverges near the  singularity as expected.}
\end{figure}

The behaviour of $T_{-}$ has been plotted against the conformal time 
$\eta$ for the detector in \fig{Paper3:FigRadInTemp01}. The effective temperature remains finite 
at the horizon crossing $\eta _{H}\sim 2.82$, while diverges later as the detector hits the 
singularity. We have also plotted the behavior of $T_{-}^{\textrm{Asym}}$ and $T_{-}^{\textrm{Hor}}$ 
as a function of the energy of the in-falling detector. We note that for unbound trajectories 
$(E \geq 1)$, the horizon crossing temperature is always greater than the 
asymptotic value.

\section{Conclusion}
\label{Paper3:SecConc}

While the quantum field theory  outside  the black hole event horizon is well studied in the literature, the corresponding issues in the region inside the event horizon have not attracted sufficient attention. This was the key motivation for this paper. We have considered a massless scalar field in the background geometry of a collapsing dust ball (Oppenheimer-Snyder model) in the in-vacuum state defined on $\mathcal{J}^-$. We focussed on the (regularised) stress-energy tensor in this vacuum state as a physically relevant diagnostic of the quantum effects. The stress-energy tensor, in turn, leads to two scalar observables -- energy density and flux in the normal direction -- related to an observer at any point in the spacetime. The key results obtained through this analysis are summarized below:

\begin{itemize}

\item Let us first start with the radially in-falling observers \emph{inside} the dust sphere and the invariant observables they measure along their trajectory.

\begin{enumerate}

\item In (1+1) spacetime dimensions the radially in-falling observers \emph{inside} the dust sphere find that the corresponding energy density and flux becomes arbitrarily large as they approach the singularity  (arbitrarily close) in a finite proper time. Even though the energy density of the dust ball ($\rho$) itself diverges as the singularity is approached, scalar field energy density ($\mathcal{U}$) diverges faster. This can be seen from the ratio $\mathcal{U}/\rho$, which diverges as $\epsilon ^{-2}$. The same conclusion holds for the total energy within an infinitesimal volume as well. Hence in (1+1) spacetime dimensions the scalar field dominates over the classical source. 

\item In (1+3) spacetime dimensions also the energy densities of both   the scalar field and classical matter becomes arbitrarily large  close to the singularity. Just as in (1+1), in (1+3) as well the scalar field energy density dominates over  the classical source.  When integrated over a small volume near the singularity energy in the scalar field goes as $\epsilon$, while that in the classical matter goes as $\epsilon ^{3}$. So, even though both of them vanish, in the $\epsilon \rightarrow 0$ limit, the classical source vanishes more rapidly. Thus even in terms of the energy the scalar field dominates over the energy in classical background. Thus backreaction is \emph{important} inside the dust sphere.

\end{enumerate}

\item The second class of observers we consider are the radially in-falling observers \emph{outside} the dust sphere. For them, we arrive at the following conclusions for the invariant observables:

\begin{enumerate}

\item In (1+1) spacetime dimensions,  radially in-falling observers \emph{outside} the dust sphere find that  the energy density of the scalar field becomes arbitrarily large  near the singularity. The same is also true for fluxes measured by these radially in-falling observers near the singularity. This  conclusion  holds even after integration over a small volume around the singularity. Since there is no classical matter present in this region this divergent energy of the scalar field has the potential to alter the singularity structure.

\item The corresponding situation in the (1+3) spacetime dimensions is similar and both the energy density and flux diverge as the singularity is approached. Even after incorporating the effect of shrinking volume  element $d^{3}x \sqrt{h}$, the divergence  persists, but becomes \emph{milder} ($\epsilon^{-4/3}$), compared to the divergence ($\epsilon^{-2}$) in (1+1) spacetime dimensions. The divergence in the energy density of the scalar field can \emph{prohibit} the formation of singularity itself due to existence of  back-reaction near the singularity. 

\end{enumerate}

Thus, the energy density in the scalar field dominates over the classical background energy density making the backreaction  important both inside and outside the dust sphere. All these conclusions hold for energy density and flux as measured along the events in the null rays hitting the singularity either inside the dust sphere or outside.

\item The radially in-falling observers outside the dust sphere observe that there is nothing peculiar at the event horizon; neither the energy density nor the flux diverges there. 

\item The study of the events along the null ray straddling on the outside of the event horizon leads to the standard Hawking energy density and flux at late times, while the events along the ray inside the horizon sees an increase of both energy density and flux to arbitrarily high values as it approaches near the singularity. Even though derived in (1+1) spacetime dimensions, the same results hold in (1+3) spacetime dimensions as well, with milder divergences. 

\end{itemize}

In addition to the above main results, we have also obtained several results for the observers on the outside of the event horizon which agree with the previous studies and expectations. Also, for completeness, we have studied the effective temperature formalism as well and observed similar effects of divergence at the singularity in this case as well.

\section*{Acknowledgements}

Work of S.C. is funded by SPM fellowship from CSIR, Govt. of India. S.S. is supported by Dr. D. S. Kothari Postdoctoral Fellowship from University Grants Commission, Govt. of India. Research of T.P. is partially supported by J.C. Bose research grant of DST, Govt. of India. Authors thank Kinjalk Lochan, Krishnamohan Parattu, Aseem Paranjape and Mandar Patil for helpful discussions. We also thank the referee for useful comments and for bringing references \cite{Bardeen2014, Hiscock1997} to our attention.

\appendix

\section{Appendix: Details of the calculations}

In this Appendix, we present the steps for deriving various 
results mentioned in the text.

\subsection{Stress-Energy Tensor: Explicit Derivation}
\label{Paper3:AppEMT}

\subsubsection{Exterior Region}\label{Paper3:AppEMTExt}

The various derivatives of the conformal factor in the exterior region have the following 
expressions:
\be
\partial _{+}C=\frac{r-1}{2r^{3}}\frac{dB/dU}{dA/dU};\hspace{15pt}\partial _{+}^{2}C=\frac{(r-1)(3-2r)}{4r^{5}}\frac{dB/dU}{dA/dU}
\ee
\be
\partial _{-}C=\frac{r-1}{r}\partial _{-}\left(\frac{dB/dU}{dA/dU}\right)-\frac{r-1}{2r^{3}}
\left(\frac{dB/dU}{dA/dU} \right)^{2}
\ee
\be
\partial _{-}^{2}C=-\frac{3}{2}\frac{r-1}{r^{3}}\frac{dB/dU}{dA/dU}\partial _{-}
\left(\frac{dB/dU}{dA/dU} \right)+\frac{r-1}{r}\partial _{-}^{2}\left(\frac{dB/dU}{dA/dU}\right)
+\frac{(3-2r)(r-1)}{4r^{5}}\left(\frac{dB/dU}{dA/dU}\right)^{3}
\ee
\be
\partial _{-}\partial _{+}C=\frac{r-1}{2r^{3}}\partial _{-}\left(\frac{dB/dU}{dA/dU}\right)
-\left(\frac{dB/dU}{dA/dU}\right)^{2}\frac{(3-2r)(r-1)}{4r^{5}}
\ee
With the following expressions for energy momentum tensor:
\be
\langle T_{++}\rangle =\frac{\kappa ^{2}}{48\pi}\left(\frac{3}{r^{4}}-\frac{4}{r^{3}}\right);\hspace{15pt}\langle T_{+-}\rangle =\frac{\kappa ^{2}}{12\pi}\frac{r-1}{r^{4}}\frac{dB/dU}{dA/dU}
\ee
\begin{align}
\langle T_{--}\rangle &= \frac{\kappa ^{2}}{48\pi}\left[\left(\frac{dB/dU}{dA/dU}\right)^{2}
\left(\frac{3}{r^{4}}-\frac{4}{r^{3}}\right)
+16\left\lbrace \frac{1}{2}\frac{\partial _{-}^{2}\left(\frac{dB/dU}{dA/dU}\right)}
{\frac{dB/dU}{dA/dU}}-\frac{3}{4}\left(\frac{\partial _{-}\left(\frac{dB/dU}{dA/dU}\right)}{\frac{dB/dU}{dA/dU}} \right)^{2} \right\rbrace \right]
\nonumber
\\
&=\frac{\kappa ^{2}}{48\pi}\left(\frac{dB/dU}{dA/dU}\right)^{2}
\Big[\left(\frac{3}{r^{4}}-\frac{4}{r^{3}}\right)+\frac{16}{\left(\frac{dB}{dU}\right)^{2}}
\nonumber
\\
&\times \left[\left\lbrace 
\frac{1}{2}\frac{\partial _{U}^{2}\left(dB/dU\right)}{dB/dU}
-\frac{3}{4}\left(\frac{\partial _{U}\left(dB/dU\right)}{dB/dU} \right)^{2} \right\rbrace
-\left\lbrace 
\frac{1}{2}\frac{\partial _{U}^{2}\left(dA/dU\right)}{dA/dU}
-\frac{3}{4}\left(\frac{\partial _{U}\left(dA/dU\right)}{dA/dU} \right)^{2} \right\rbrace
 \right]\Big]
\end{align}
these relations can be simplified to arrive at,
\be
\langle T_{+-}\rangle =\frac{\kappa ^{2}}{12\pi}\frac{r-1}{r^{4}}
\left(\frac{\cot \chi _{0}+\tan \left(\frac{U-\chi _{0}}{2}\right)}
{\cot \chi _{0}-\tan \left(\frac{U+\chi _{0}}{2}\right)}\right)
\frac{\cos ^{2}\left(\frac{U+\chi _{0}}{2}\right)}{\cos ^{2}\left(\frac{U-\chi _{0}}{2}\right)}
\ee

\begin{align}
\langle T_{--}\rangle &=\frac{\kappa ^{2}}{48\pi}
\left(\left(\frac{\cot \chi _{0}+\tan \left(\frac{U+\chi _{0}}{2}\right)}
{\cot \chi _{0}-\tan \left(\frac{U+\chi_{0}}{2}\right)}\right)
\frac{\cos ^{2}\left(\frac{U+\chi _{0}}{2}\right)}{\cos ^{2}\left(\frac{U-\chi _{0}}{2}\right)}
\right)^2 \left[\left(\frac{3}{r^{4}}-\frac{4}{r^{3}}\right)\right.\nonumber\\
&+\left(\frac{a_{max}\cos ^{2}\left(\frac{U+\chi _{0}}{2}\right)}
{\sin \chi _{0}\left(\cot \chi _{0}-\tan \left(\frac{U+\chi _{0}}{2}\right)\right)}\right)^{-2}
\left(-15\left[\tan ^{2}\left(\frac{U+\chi _{0}}{2}\right)-\tan ^{2}\left(\frac{U-\chi _{0}}{2}\right)\right]\right.\nonumber\\
&-6\cot \chi_{0}\left[\tan \left(\frac{U+\chi _{0}}{2}\right) +\tan\left(\frac{U-\chi _{0}}{2}\right)\right]
+\frac{4\cot \chi _{0}}{\sin ^{2}\chi _{0}}\left[\frac{1}{\cot \chi _{0}-\tan \left(\frac{U+\chi _{0}}{2}\right)}\right.\nonumber\\
&\left.-\frac{1}{\cot \chi _{0}+\tan \left(\frac{U-\chi _{0}}{2}\right)}\right]
+\frac{a_{max}}{\sin \chi _{0}}\left[\frac{1}{\left(\cot \chi _{0} -\tan \left(\frac{U+\chi _{0}}{2}\right)\right)^{2}}\right.
\nonumber\\
&-\left.\left.\left.\frac{1}{\left(\cot \chi _{0}+\tan \left(\frac{U-\chi _{0}}{2}\right)\right)^{2}} \right]
\right)\right]
\end{align}
In arriving at the above relations we have 
used the following expressions for the various 
derivatives $dA/dU$ and $dB/dU$ are respectively:
\be
\frac{dA}{dU}=\frac{a_{max}\cos ^{2}\left(\frac{U-\chi _{0}}{2}\right)}
{\sin \chi _{0}\left(\cot \chi _{0}+\tan \left(\frac{U-\chi _{0}}{2}\right)\right)};\hspace{15pt}\frac{dB}{dU}=\frac{a_{max}\cos ^{2}\left(\frac{U+\chi _{0}}{2}\right)}
{\sin \chi _{0}\left(\cot \chi _{0}-\tan \left(\frac{U+\chi _{0}}{2}\right)\right)}
\ee

\be
\frac{d^{2}A}{dU^{2}}=-\frac{1}{2}\frac{\left[1+2\sin ^{2}\left(\frac{U-\chi _{0}}{2}\right)
+\cot \chi _{0}\sin \left(U-\chi _{0}\right)\right]}{\sin ^{4}\chi _{0}\left(\cot \chi _{0}+\tan \left(\frac{U-\chi _{0}}{2}\right)\right)^{2}}
\ee
\be
\frac{d^{2}B}{dU^{2}}=\frac{1}{2}\frac{\left[1+2\sin ^{2}\left(\frac{U+\chi _{0}}{2}\right)
-\cot \chi _{0}\sin \left(U+\chi _{0}\right)\right]}{\sin ^{4}\chi _{0}\left(\cot \chi _{0}-\tan \left(\frac{U+\chi _{0}}{2}\right)\right)^{2}}
\ee
as well as the following derivatives:
\begin{align}
\frac{1}{2}\frac{\partial _{U}^{2}\left(dA/dU\right)}{dA/dU}
&-\frac{3}{4}\left(\frac{\partial _{U}\left(dA/dU\right)}{dA/dU} \right)^{2}
= -\frac{1}{16}\Big[15\tan ^{4}\left(\frac{U-\chi _{0}}{2}\right)+24\cot \chi _{0}
\tan ^{3}\left(\frac{U-\chi _{0}}{2}\right)
\nonumber
\\
&+10\tan ^{2} \left(\frac{U-\chi _{0}}{2} \right)
+4\cot ^{2}\chi _{0}\left\lbrace 1+2\tan ^{2}\left(\frac{U-\chi _{0}}{2}\right)\right\rbrace
\nonumber
\\
&+16\cot \chi _{0}\tan \left(\frac{U-\chi _{0}}{2}\right)-1\Big]
\left[\cot \chi _{0}+\tan \left(\frac{U-\chi _{0}}{2}\right) \right]^{-2}
\nonumber
\\
&=-\frac{1}{16}\Big[15\tan ^{2}\left(\frac{U-\chi _{0}}{2}\right)
-6\cot \chi _{0}\tan \left(\frac{U-\chi _{0}}{2}\right)+5\left(1+\sin ^{-2}\chi _{0}\right)
\nonumber
\\
&-\frac{4\sin ^{-2}\chi _{0}\cot \chi _{0}}{\cot \chi _{0}+\tan \left(\frac{U-\chi _{0}}{2}\right)}
-\frac{a_{max}}{\sin \chi _{0}\left(\cot \chi _{0}
+\tan \left(\frac{U-\chi _{0}}{2}\right) \right)^{2}}\Big]
\end{align}
\begin{align}
\frac{1}{2}\frac{\partial _{U}^{2}\left(dB/dU\right)}{dB/dU}
&-\frac{3}{4}\left(\frac{\partial _{U}\left(dB/dU\right)}{dB/dU} \right)^{2}
= -\frac{1}{16}\Big[15\tan ^{4}\left(\frac{U+\chi _{0}}{2}\right)-24\cot \chi _{0}
\tan ^{3}\left(\frac{U+\chi _{0}}{2}\right)
\nonumber
\\
&-10\tan ^{2}\left(\frac{U+\chi _{0}}{2} \right)
+4\cot ^{2}\chi _{0}\left\lbrace 1+2\tan ^{2}\left(\frac{U+\chi _{0}}{2}\right)\right\rbrace
\nonumber
\\
&-16\cot \chi _{0}\tan \left(\frac{U+\chi _{0}}{2}\right)-1\Big]
\left[\cot \chi _{0}-\tan \left(\frac{U+\chi _{0}}{2}\right) \right]^{-2}
\nonumber
\\
&=-\frac{1}{16}\Big[15\tan ^{2}\left(\frac{U+\chi _{0}}{2}\right)
+6\cot \chi _{0}\tan \left(\frac{U+\chi _{0}}{2}\right)+5\left(1+\sin ^{-2}\chi _{0}\right)
\nonumber
\\
&-\frac{4\sin ^{-2}\chi _{0}\cot \chi _{0}}{\cot \chi _{0}-\tan \left(\frac{U+\chi _{0}}{2}\right)}
-\frac{a_{max}}{\sin \chi _{0}\left(\cot \chi _{0}
-\tan \left(\frac{U+\chi _{0}}{2}\right) \right)^{2}}\Big]
\end{align}

\subsubsection{Interior Region}\label{Paper3:AppEMTInt}

For the interior region the various derivatives of the conformal factors are:
\be
\frac{1}{C}\partial _{+}C=\frac{1}{\left(dA/dV\right)}\left[\frac{1}{a^{2}}\frac{da^{2}}{dV}-
\frac{d^{2}A/dV^{2}}{dA/dV}\right];\hspace{15pt}\frac{1}{C}\partial _{-}C=\frac{1}{\left(dA/dU\right)}\left[\frac{1}{a^{2}}\frac{da^{2}}{dU}-
\frac{d^{2}A/dU^{2}}{dA/dU}\right]
\ee

\be
\frac{1}{C}\partial _{+}^{2}C=\frac{1}{\left(dA/dV\right)^{2}}
\left[\frac{1}{a^{2}}\frac{d^{2}a^{2}}{dV^{2}}
-\frac{3}{a^{2}}\frac{da^{2}}{dV}\frac{1}{dA/dV}\frac{d^{2}A}{dV^{2}}
-\frac{1}{dA/dV}\frac{d^{3}A}{dV^{3}}
+3\left(\frac{1}{dA/dV}\frac{d^{2}A}{dV^{2}}\right)^{2} \right]
\ee

\be
\frac{1}{C}\partial _{-}^{2}C=\frac{1}{\left(dA/dU\right)^{2}}
\left[\frac{1}{a^{2}}\frac{d^{2}a^{2}}{dU^{2}}
-\frac{3}{a^{2}}\frac{da^{2}}{dU}\frac{1}{dA/dU}\frac{d^{2}A}{dU^{2}}
-\frac{1}{dA/dU}\frac{d^{3}A}{dU^{3}}
+3\left(\frac{1}{dA/dU}\frac{d^{2}A}{dU^{2}}\right)^{2} \right]
\ee

\begin{align}
\frac{1}{C}\partial _{-}\partial _{+}C&=\frac{1}{\frac{dA}{dV}\frac{dA}{dU}}
\left[\frac{1}{a^{2}}\frac{d^{2}a^{2}}{dUdV}
-\frac{1}{a^{2}}\frac{da^{2}}{dU}\frac{1}{dA/dV}\frac{d^{2}A}{dV^{2}}
-\frac{1}{a^{2}}\frac{da^{2}}{dV}\frac{1}{dA/dU}\frac{d^{2}A}{dU^{2}}\right.\nonumber\\
&\left.+\frac{1}{dA/dV}\frac{d^{2}A}{dV^{2}}\frac{1}{dA/dU}\frac{d^{2}A}{dU^{2}}\right]
\end{align}
Then the components of the stress energy tensor in the inside region are:
\begin{align}
\langle T_{++}\rangle &=\frac{1}{12\pi}\frac{1}{\left(dA/dV\right)^{2}}
\left[\left\lbrace\frac{1}{2a^{2}}\frac{d^{2}a^{2}}{dV^{2}}
-\frac{3}{4}\left(\frac{1}{a^{2}}\frac{da^{2}}{dV}\right)^{2} \right\rbrace 
-\left\lbrace \frac{1}{2}\frac{1}{dA/dV}\frac{d^{3}A}{dV^{3}}\right.\right.\nonumber\\
&\left.\left.-\frac{3}{4}\frac{1}{\left(dA/dV\right)^{2}}\left(\frac{d^{2}A}{dV^{2}}\right)^{2} \right\rbrace \right]
\nonumber
\\
&=\frac{1}{12\pi}\frac{1}{\left(dA/dV\right)^{2}}\left[-\frac{1}{8}
\left(\frac{3a_{max}\sin \chi _{0}}{r\left(\frac{U+V}{2}\right)}-2\right)
-\left\lbrace \frac{1}{2}\frac{\partial _{V}^{2}\left(dA/dV\right)}{\left(dA/dV\right)}
-\frac{3}{4}\left(\frac{\partial _{V}\left(dA/dV\right)}{\left(dA/dV\right)}\right)^{2} \right\rbrace  \right]
\end{align}
\begin{align}
\langle T_{--}\rangle &=\frac{1}{12\pi}\frac{1}{\left(dA/dU\right)^{2}}
\left[\left\lbrace\frac{1}{2a^{2}}\frac{d^{2}a^{2}}{dU^{2}}
-\frac{3}{4}\left(\frac{1}{a^{2}}\frac{da^{2}}{dU}\right)^{2} \right\rbrace 
-\left\lbrace \frac{1}{2}\frac{1}{dA/dU}\frac{d^{3}A}{dU^{3}}\right.\right.\nonumber\\
&\left.\left.-\frac{3}{4}\frac{1}{\left(dA/dU\right)^{2}}\left(\frac{d^{2}A}{dU^{2}}\right)^{2} \right\rbrace \right]
\nonumber
\\
&=\frac{1}{12\pi}\frac{1}{\left(dA/dU\right)^{2}}
\left[-\frac{1}{8}
\left(\frac{3a_{max}\sin \chi _{0}}{r\left(\frac{U+V}{2}\right)}-2\right)
-\left\lbrace \frac{1}{2}\frac{\partial _{U}^{2}\left(dA/dU\right)}{\left(dA/dU\right)}\right.\right.\nonumber\\
&\left.\left.-\frac{3}{4}\left(\frac{\partial _{U}\left(dA/dU\right)}{\left(dA/dU\right)}\right)^{2}  \right\rbrace \right]
\end{align}
\begin{align}
\langle T_{+-}\rangle &=\frac{1}{24\pi}\frac{1}{dA/dV}\frac{1}{dA/dU}
\left[\frac{1}{a^{2}}\frac{d^{2}a^{2}}{dUdV}
-\frac{1}{a^{2}}\frac{da^{2}}{dU}\frac{1}{a^{2}}\frac{da^{2}}{dV}\right]
\nonumber
\\
&=-\frac{1}{48\pi}\frac{1}{dA/dV}\frac{1}{dA/dU}\frac{1}{1+\cos \left(\frac{U+V}{2} \right)}
\end{align}
where various derivatives of the quantity $A$ are given in 
\sect{Paper3:AppEMTExt}.

\subsection{Energy Density and Flux For Various Observers}
\label{Paper3:AppENGFLUX}

\subsubsection{Static Observer}
\label{Paper3:AppENGFLUXStatic}

Below we provide the full expressions for energy density and flux 
calculated for static observer:

\begin{align}
\mathcal{U}&=\langle T_{++}\rangle \left(\dot{V}^{+}\right)^{2}
+\langle T_{--}\rangle \left(\dot{V}^{-}\right)^{2}
+2\langle T_{+-}\rangle \dot{V}^{+}\dot{V}^{-}
\nonumber
\\
&=\frac{\kappa ^{2}}{48\pi}\left(\frac{r}{r-1}\right)
\Big[\left(-\frac{2}{r^{4}}\right)+\frac{\sin ^{2}\chi _{0}}{a_{max}^{2}\cos ^{4}
\left(\frac{U+\chi _{0}}{2}\right)}\left(\cot \chi _{0}-\tan \left(\frac{U+\chi _{0}}{2}\right)\right)^{2}
\nonumber
\\
&\times \Big\lbrace 
-15\left[\tan ^{2}\left(\frac{U+\chi _{0}}{2}\right)-\tan ^{2}\left(\frac{U-\chi _{0}}{2}\right) \right]-6\cot \chi _{0}\left[\tan \left(\frac{U+\chi _{0}}{2}\right)
+\tan\left(\frac{U-\chi _{0}}{2}\right)\right]
\nonumber
\\
&+\frac{4\cot \chi _{0}}{\sin ^{2}\chi _{0}}\left[\frac{1}{\cot \chi _{0}-\tan \left(\frac{U+\chi _{0}}{2}\right)}-\frac{1}{\cot \chi _{0}+\tan \left(\frac{U-\chi _{0}}{2}\right)}\right]
\Big\rbrace 
\nonumber
\\
&+\frac{\sin \chi _{0}}
{a_{max}\cos ^{4}\left(\frac{U+\chi _{0}}{2}\right)}\left[1
-\frac{\left(\cot \chi _{0}-\tan \left(\frac{U+\chi _{0}}{2}\right)\right)^{2}}{\left(\cot \chi _{0}+\tan \left(\frac{U-\chi _{0}}{2}\right)\right)^{2}} \right]
\Big]
\end{align}
and the expression for flux turns out to be:
\begin{align}
\mathcal{F}&=-\langle T_{ab}\rangle u^{a}n^{b}=-\langle T_{++}\rangle \left(\dot{V}^{+}\right)^{2}
+\langle T_{--}\rangle \left(\dot{V}^{-}\right)^{2}
\nonumber
\\
&=\frac{\kappa ^{2}}{48\pi}\left(\frac{r}{r-1}\right)
\Big[\frac{\sin ^{2}\chi _{0}}{a_{max}^{2}\cos ^{4}
\left(\frac{U+\chi _{0}}{2}\right)}\left(\cot \chi _{0}-\tan \left(\frac{U+\chi _{0}}{2}\right)\right)^{2}
\nonumber
\\
&\times \Big\lbrace 
-15\left[\tan ^{2}\left(\frac{U+\chi _{0}}{2}\right)-\tan ^{2}\left(\frac{U-\chi _{0}}{2}\right) \right]-6\cot \chi _{0}\left[\tan \left(\frac{U+\chi _{0}}{2}\right)
+\tan\left(\frac{U-\chi _{0}}{2}\right)\right]
\nonumber
\\
&+\frac{4\cot \chi _{0}}{\sin ^{2}\chi _{0}}\left[\frac{1}{\cot \chi _{0}-\tan \left(\frac{U+\chi _{0}}{2}\right)}-\frac{1}{\cot \chi _{0}+\tan \left(\frac{U-\chi _{0}}{2}\right)}\right]
\Big\rbrace 
\nonumber
\\
&+\frac{\sin \chi _{0}}
{a_{max}\cos ^{4}\left(\frac{U+\chi _{0}}{2}\right)}\left[1
-\frac{\left(\cot \chi _{0}-\tan \left(\frac{U+\chi _{0}}{2}\right)\right)^{2}}{\left(\cot \chi _{0}+\tan \left(\frac{U-\chi _{0}}{2}\right)\right)^{2}} \right]
\Big]
\end{align}

\subsubsection{Radially In-falling Observers: Inside}
\label{Paper3:AppENGFluxRadialIn}

The energy density for radially in-falling observer 
has the following expression:

\begin{align}\label{Coveq02}
\mathcal{U}&=\frac{\kappa ^{2}}{48\pi}\frac{1}{a^{2}\left(\eta \right)}\Big[-8
\sec ^{2}\frac{\eta}{2}
+4+\frac{1}{2}\Big \lbrace 15\tan ^{2}\left(\frac{\eta -\chi _{0}-\tilde{\chi}}{2}\right)
-6\cot \chi _{0}\tan \left(\frac{\eta -\chi _{0}-\tilde{\chi}}{2}\right)+5\left(1+\sin ^{-2}\chi _{0}\right)
\nonumber
\\
&-\frac{4\sin ^{-2}\chi _{0}\cot \chi _{0}}{\cot \chi _{0}
+\tan \left(\frac{\eta -\chi _{0}-\tilde{\chi}}{2}\right)}
-\frac{a_{max}}{\sin \chi _{0}\left(\cot \chi _{0}
+\tan \left(\frac{\eta -\chi _{0}-\tilde{\chi}}{2}\right) \right)^{2}}
+15\tan ^{2}\left(\frac{\eta -\chi _{0}+\tilde{\chi}}{2}\right)
\nonumber
\\
&-6\cot \chi _{0}\tan \left(\frac{\eta -\chi _{0}+\tilde{\chi}}{2}\right)+5\left(1+\sin ^{-2}\chi _{0}\right)
-\frac{4\sin ^{-2}\chi _{0}\cot \chi _{0}}{\cot \chi _{0}
+\tan \left(\frac{\eta -\chi _{0}+\tilde{\chi}}{2}\right)}
\nonumber
\\
&-\frac{a_{max}}{\sin \chi _{0}\left(\cot \chi _{0}
+\tan \left(\frac{\eta -\chi _{0}+\tilde{\chi}}{2}\right) \right)^{2}}
\Big\rbrace \Big]
\end{align}
while the flux has the following expression:
\begin{align}
\mathcal{F}&=\frac{\kappa ^{2}}{48\pi}\frac{1}{a^{2}\left(\eta \right)}\frac{1}{2}
\Big \lbrace 15\tan ^{2}\left(\frac{\eta -\chi _{0}-\tilde{\chi}}{2}\right)
-6\cot \chi _{0}\tan \left(\frac{\eta -\chi _{0}-\tilde{\chi}}{2}\right)+5\left(1+\sin ^{-2}\chi _{0}\right)
\nonumber
\\
&-\frac{4\sin ^{-2}\chi _{0}\cot \chi _{0}}{\cot \chi _{0}
+\tan \left(\frac{\eta -\chi _{0}-\tilde{\chi}}{2}\right)}
-\frac{a_{max}}{\sin \chi _{0}\left(\cot \chi _{0}
+\tan \left(\frac{\eta -\chi _{0}-\tilde{\chi}}{2}\right) \right)^{2}}
-15\tan ^{2}\left(\frac{\eta -\chi _{0}+\tilde{\chi}}{2}\right)
\nonumber
\\
&+6\cot \chi _{0}\tan \left(\frac{\eta -\chi _{0}+\tilde{\chi}}{2}\right)-5\left(1+\sin ^{-2}\chi _{0}\right)
+\frac{4\sin ^{-2}\chi _{0}\cot \chi _{0}}{\cot \chi _{0}
-\tan \left(\frac{\eta -\chi _{0}+\tilde{\chi}}{2}\right)}
\nonumber
\\
&+\frac{a_{max}}{\sin \chi _{0}\left(\cot \chi _{0}
+\tan \left(\frac{\eta -\chi _{0}+\tilde{\chi}}{2}\right) \right)^{2}}
\Big\rbrace 
\end{align}
\subsubsection{Radially In-falling Observers: Outside}
\label{Paper3:AppENGFluxRadialOut}

For radially in-falling observer outside the dust ball 
has the following expression for energy density:
\begin{align}\label{EngAppTotal}
\mathcal{U}&=\langle T_{++}\rangle \left(\dot{V}^{+}\right)^{2}
+\langle T_{--}\rangle \left(\dot{V}^{-}\right)^{2}
+2\langle T_{+-}\rangle \dot{V}^{+}\dot{V}^{-}
\nonumber
\\
&=\frac{\kappa ^{2}}{48\pi} 4E^{2}\left(\frac{r}{r-1}\right)^{2}
\left(\frac{3}{r^{4}}-\frac{4}{r^{3}}\right)
+\frac{\kappa ^{2}}{24\pi}\left(\frac{r}{r-1}\right)
\left(-\frac{7}{r^{4}}+\frac{8}{r^{3}}\right)
\nonumber
\\
&+\frac{\kappa ^{2}}{48\pi}\left(\frac{r}{r-1}\right)^{2}
\left(E+\sqrt{E^{2}-\frac{r-1}{r}}\right)^{2}
\times \Big[\frac{\sin ^{2}\chi _{0}}{a_{max}^{2}\cos ^{4}
\left(\frac{\eta +\chi _{0}-\tilde{\chi}}{2}\right)}\left(\cot \chi _{0}-\tan \left(\frac{\eta +\chi _{0}-\tilde{\chi}}{2}\right)\right)^{2}
\nonumber
\\
&\times \Big\lbrace 
-15\left[\tan ^{2}\left(\frac{\eta +\chi _{0}-\tilde{\chi}}{2}\right)-\tan ^{2}\left(\frac{\eta -\chi _{0}-\tilde{\chi}}{2}\right) \right]
-6\cot \chi _{0}\left[\tan \left(\frac{\eta +\chi _{0}-\tilde{\chi}}{2}\right)
+\tan\left(\frac{\eta -\chi _{0}-\tilde{\chi}}{2}\right)\right]
\nonumber
\\
&+\frac{4\cot \chi _{0}}{\sin ^{2}\chi _{0}}\left[\frac{1}{\cot \chi _{0}-\tan \left(\frac{\eta +\chi _{0}-\tilde{\chi}}{2}\right)}-\frac{1}{\cot \chi _{0}+\tan \left(\frac{\eta -\chi _{0}-\tilde{\chi}}{2}\right)}\right]
\Big\rbrace 
\nonumber
\\
&+\frac{\sin \chi _{0}}
{a_{max}\cos ^{4}\left(\frac{\eta +\chi _{0}-\tilde{\chi}}{2}\right)}\left[1
-\frac{\left(\cot \chi _{0}-\tan \left(\frac{\eta +\chi _{0}-\tilde{\chi}}{2}\right)\right)^{2}}{\left(\cot \chi _{0}+\tan \left(\frac{\eta -\chi _{0}-\tilde{\chi}}{2}\right)\right)^{2}} \right]
\Big]
\end{align}
and the flux has the following expression:
\begin{align}
\mathcal{F}&=-\langle T_{ab}\rangle u^{a}n^{b}=-\langle T_{++}\rangle \left(\dot{V}^{+}\right)^{2}
+\langle T_{--}\rangle \left(\dot{V}^{-}\right)^{2}
\nonumber
\\
&=\frac{\kappa ^{2}}{48 \pi}4E\sqrt{E^{2}-\frac{r-1}{r}} \left(\frac{r}{r-1}\right)^{2}
\left(\frac{3}{r^{4}}-\frac{4}{r^{3}}\right)
+\frac{\kappa ^{2}}{48\pi}\left(\frac{r}{r-1}\right)^{2}
\left(E+\sqrt{E^{2}-\frac{r-1}{r}}\right)^{2}
\nonumber
\\
&\times \Big[\frac{\sin ^{2}\chi _{0}}{a_{max}^{2}\cos ^{4}
\left(\frac{\eta +\chi _{0}-\tilde{\chi}}{2}\right)}\left(\cot \chi _{0}-\tan \left(\frac{\eta +\chi _{0}-\tilde{\chi}}{2}\right)\right)^{2}
\nonumber
\\
&\times \Big\lbrace 
-15\left[\tan ^{2}\left(\frac{\eta +\chi _{0}-\tilde{\chi}}{2}\right)-\tan ^{2}\left(\frac{\eta -\chi _{0}-\tilde{\chi}}{2}\right) \right]
\nonumber
\\
&-6\cot \chi _{0}\left[\tan \left(\frac{\eta +\chi _{0}-\tilde{\chi}}{2}\right)
+\tan\left(\frac{\eta -\chi _{0}-\tilde{\chi}}{2}\right)\right]
\nonumber
\\
&+\frac{4\cot \chi _{0}}{\sin ^{2}\chi _{0}}\left[\frac{1}{\cot \chi _{0}-\tan \left(\frac{\eta +\chi _{0}-\tilde{\chi}}{2}\right)}-\frac{1}{\cot \chi _{0}+\tan \left(\frac{\eta -\chi _{0}-\tilde{\chi}}{2}\right)}\right]
\Big\rbrace 
\nonumber
\\
&+\frac{\sin \chi _{0}}
{a_{max}\cos ^{4}\left(\frac{\eta +\chi _{0}-\tilde{\chi}}{2}\right)}\left[1
-\frac{\left(\cot \chi _{0}-\tan \left(\frac{\eta +\chi _{0}-\tilde{\chi}}{2}\right)\right)^{2}}{\left(\cot \chi _{0}+\tan \left(\frac{\eta -\chi _{0}-\tilde{\chi}}{2}\right)\right)^{2}} \right]
\Big]
\end{align}

\subsection{Effective temperature for various observers}

In this section we present the full expression for effective temperature measured by detectors in various trajectories. For static observer the effective temperature turns out to be:
\begin{align}\label{Paper3:Eq29}
T_{-}&=\frac{1}{2\pi} \left \vert \frac{\ddot{V}^{-}}{\dot{V}^{-}}\right\vert 
\nonumber
\\
&=\frac{1}{2\pi}\left \vert 
\dot{u}~\left(\frac{\frac{d^{2}A}{dU^{2}}}{\frac{dA}{dU}\frac{dB}{dU}}
-\frac{\frac{d^{2}B}{dU^{2}}}{\left(\frac{dB}{dU}\right)^{2}} \right)\right \vert
\nonumber
\\
&=\frac{1}{4\pi}\sqrt{\frac{r}{r-1}}\sin ^{4}\chi _{0}
\Big\lbrace \frac{\left[1+2\sin ^{2}\left(\frac{U-\chi _{0}}{2}\right)
+\cot \chi _{0}\sin \left(U-\chi _{0}\right)\right]
\left(\cot \chi _{0}-\tan \left(\frac{U+\chi _{0}}{2}\right)\right)}
{\left(\cot \chi _{0}+\tan \left(\frac{U-\chi _{0}}{2}\right)\right)
\cos ^{2}\left(\frac{U+\chi _{0}}{2}\right)
\cos ^{2}\left(\frac{U-\chi _{0}}{2}\right)} 
\nonumber
\\
&+\frac{\left[1+2\sin ^{2}\left(\frac{U+\chi _{0}}{2}\right)
-\cot \chi _{0}\sin \left(U+\chi _{0}\right)\right]}{\cos ^{4}\left(\frac{U+\chi _{0}}{2}\right)}
\Big\rbrace 
\end{align}
For radially in-falling observes inside the dust sphere the effective temperature turns out to be:
\begin{equation}\label{Paper3:Eq32}
T_{-}=\frac{1}{4\pi}\left \vert \frac{1
+2\sin ^{2}\left(\frac{\eta -\chi _{0}-\tilde{\chi}}{2}\right)
+\cot \chi _{0}\sin \left(\frac{\eta -\chi _{0}-\tilde{\chi}}{2}\right)}
{\cos ^{2}\left(\frac{\eta -\chi _{0}-\tilde{\chi}}{2}\right)
\left\lbrace \cot \chi _{0}+\tan \left(\frac{\eta -\chi _{0}-\tilde{\chi}}{2}\right)
\right\rbrace} \right \vert
\end{equation}
Finally, for radially in-falling observers in the Schwarzschild spacetime effective temperature takes the following expression:
\begin{align}\label{Paper3:Eq41}
T_{-}&=\frac{1}{2\pi}\Big \vert \frac{\ddot{u}}{\dot{u}}
-\dot{u}\frac{\sin ^{4}\chi _{0}}{2}
\Big\lbrace \frac{\left[1+2\sin ^{2}\left(\frac{\eta -\chi _{0}-\tilde{\chi}}{2}\right)
+\cot \chi _{0}\sin \left(\eta -\chi _{0}-\tilde{\chi}\right)\right]
\left(\cot \chi _{0}-\tan \left(\frac{\eta +\chi _{0}-\tilde{\chi}}{2}\right)\right)}
{\left(\cot \chi _{0}+\tan \left(\frac{\eta -\chi _{0}-\tilde{\chi}}{2}\right)\right)
\cos ^{2}\left(\frac{\eta +\chi _{0}-\tilde{\chi}}{2}\right)
\cos ^{2}\left(\frac{\eta -\chi _{0}-\tilde{\chi}}{2}\right)} 
\nonumber
\\
&+\frac{\left[1+2\sin ^{2}\left(\frac{\eta +\chi _{0}-\tilde{\chi}}{2}\right)
-\cot \chi _{0}\sin \left(\eta +\chi _{0}-\tilde{\chi}\right)\right]}{\cos ^{4}\left(\frac{\eta +\chi _{0}-\tilde{\chi}}{2}\right)}
\Big\rbrace 
\Big \vert 
\end{align}

\end{document}